\documentclass[12pt, centerh1]{article}

\usepackage{mathptmx,natbib,color,placeins}       % selects Times Roman as basic font
\usepackage[cal=esstix]{mathalfa}
\usepackage{amsfonts, amsmath, amssymb,marvosym,multirow,bm} %packages added later

\usepackage{makeidx}         % allows index generation
\usepackage{graphicx,subfigure}        % standard LaTeX graphics tool
\usepackage{multicol}        % used for the two-column index
\newcommand{\ex}{\mathbb{E}}

\newcommand{\tr}{\,\mbox{tr}}

\newcommand{\Beta}{\boldsymbol\beta}
\newcommand{\bbeta}{\boldsymbol\beta}
\newcommand{\ba}{\mathbf{a}}

\newcommand{\bY}{\mathbf{Y}}
\newcommand{\by}{\mathbf{y}}
\newcommand{\bz}{\mathbf{z}}

\newcommand{\Mu}{\boldsymbol{\mu}}

\newcommand{\btheta}{\boldsymbol{\theta}}

\newcommand{\bu}{\mathbf{u}}
\newcommand{\bw}{\mathbf{w}}

\title{Variational Bayes Approximations for Clustering via Mixtures of Normal Inverse Gaussian Distributions}
\author{Sanjeena Subedi\thanks{Department of Mathematics \& Statistics, University of Guelph, Guelph, Ontario, N1G 2W1, Canada. E-mail: ssubedi@uoguelph.ca.} \ and Paul D.\ McNicholas}
\date{Department of Mathematics \& Statistics, University of Guelph}

\begin{document}
\maketitle

\begin{abstract}
Parameter estimation for model-based clustering using a finite mixture of normal inverse Gaussian (NIG) distributions is achieved through variational Bayes approximations. Univariate NIG mixtures and multivariate NIG mixtures are considered. The use of variational Bayes approximations here is a substantial departure from the traditional EM approach and alleviates some of the associated computational complexities and uncertainties. Our variational algorithm is applied to simulated and real data. The paper concludes with discussion and suggestions for future work.

\end{abstract}

\section{Introduction}

The use of mixture models for clustering, referred to as model-based clustering, has become increasingly popular since the work of \cite{wolfe63}. A wide variety of finite mixture models has been studied extensively within the literature to date. Amongst these, the Gaussian mixture model has received special attention due to its mathematical tractability and the relative computational simplicity associated with parameter estimation. However, the Gaussian mixture model is not without limitations; for instance, the component densities are restricted to being symmetric. Over the past few years, there has been a notable increase in the preponderance of non-Gaussian mixture modelling within the literature \citep[e.g.,][]{lin09,lin10,andrews11b,baek11,steane12,mcnicholas12,vrbik12,browne12,morris13b,morris13a,morris13c,lee13,murray13}. \cite{karlis09} proposed a mixture of univariate normal inverse Gaussian (UNIG) distributions and a mixture of multivariate normal inverse Gaussian (MNIG) distributions; these models have the flexibility to represent both skewed and symmetric populations as well as mixtures thereof. As is typical within the field, parameter estimation for the NIG mixtures has heretofore been carried out using an expectation-maximization (EM) algorithm \citep{baum70,orchard72,sundberg74,dempster77}; see \cite{karlis09} for details.

The EM algorithm is an iterative procedure used to find maximum likelihood estimates for incomplete data. In the clustering context, the group memberships are missing and latent variables may also be present. One major drawback to the EM approach is its dependency on starting values. This and other problems arise because of the unpleasant nature of the likelihood surface, which leads to a very slow rate of convergence and, in some cases, convergence to local minima \citep[cf.][]{titterington85}. As reported by \cite{karlis09}, the EM algorithm can be very slow when dealing with complicated distributions, such as the MNIG. Furthermore, when the number of components in a mixture model is unknown, the computational cost increases further because the EM algorithm must be used in conjunction with a model-selection criterion so that every possible number of components is explored \citep[e.g.,][]{fraley02a,bouveyron07,mcnicholas08,mcnicholas10b}. Beyond the increased computational cost, this problem is further compounded by the fact that using different model selection criteria on the same data can result in selection of a different set of models \citep[see][for examples]{andrews11a}. 

Variational Bayes approximations have been explored by many researchers, including \cite{waterhouse96}, \cite{jordan99}, \cite{corduneanu01}, and \cite{mcgrory07}. Variational Bayes approximations are an iterative Bayesian alternative to the EM algorithm, and their fast and deterministic nature has made the approach increasingly popular over the past decade or so. The tractability of the variational approach allows for simultaneous model selection and parameter estimation, thus removing the need for a model selection criterion and reducing the associated computational overhead. The variational Bayes algorithm has been applied to Gaussian mixture models \citep[cf.][]{teschendorff05,mcgrory07}. For observed data $\mathbf{y}$, the joint conditional distribution of parameters $\boldsymbol{\theta}$ and missing data $\mathbf{w}$ is approximated by constructing a tight lower bound on the complex data marginal likelihood using a computationally convenient density $q_{\btheta,\bw}(\mathbf{\boldsymbol{\theta},\mathbf{w}})$. The approximating density $q_{\btheta,\bw}(\mathbf{\boldsymbol{\theta},\mathbf{w}})$ is obtained by minimizing the Kullback-Leibler (KL) divergence between the true density $h(\btheta,\mathbf{w}|\mathbf{y})$ and the approximating density \citep{beal2003,mcgrory07}. Due to the non-negative property of the KL divergence, minimizing the KL divergence is equivalent to maximizing the lower bound. The algorithm is initialized with more components than expected, and estimation of the parameters and the number of components is performed simultaneously.

In this paper, we develop a variational Bayes framework for parameter estimation for UNIG mixtures and MNIG mixtures. Using the variational Bayes framework reduces the computational cost associated with this complex modelling framework by simultaneously estimating the parameters and the number of components. We show that variational Bayes approximations can be very effective for non-Gaussian mixture model-based clustering. The remainder of this paper is laid out as follows. Variational approximations are developed and illustrated for UNIG mixture models (Section~\ref{sec:2}). They are then developed and illustrated for MNIG mixtures in Section \ref{sec:3}. The paper concludes with discussion and suggestions for future work (Section \ref{sec:4}).

\section{Mixture of Univariate Normal Inverse Gaussian Distributions}
In this section, we introduce a variational Bayes framework for parameter estimation for the UNIG mixture model.
\label{sec:2}
\subsection{The Model}
A mean-variance mixture of a univariate normal distribution with the inverse Gaussian (IG) distribution \citep{barndorff97}, i.e.,  \begin{equation*}
Y \mid u \sim \text{N}(\mu+\beta u, u), \qquad
%\end{equation*}
%\begin{equation*}
U \sim \text{IG}(\delta,\gamma),
\end{equation*}
results in a UNIG distribution with density
\begin{equation*}
f(y ; \boldsymbol{\theta})=\frac{\alpha}{\pi}\exp\left\{\delta\sqrt{\alpha^2-\beta^2}-\beta\mu\right\} \phi(y)^{-\frac{1}{2}} K_1(\delta\alpha \phi(y)^\frac{1}{2}  ) \exp\left\{\beta y\right\},
\end{equation*}
where $\boldsymbol{\theta}=(\alpha,\beta,\mu,\delta)$ are the model parameters such that $\alpha^2=\gamma^2+\beta^2$,  $\phi(y)=1+[(y-\mu)/\delta]^2$, %$\text{IG}(\cdot,\delta,\gamma)$ refers to inverse Gaussian distribution, 
and  $K_1(y)$ is the modified Bessel function of the third kind of order $1$ evaluated at $y$ \citep{abramowitz72}.  The expected value and variance of $Y$ are $\mathbb{E}(Y)=\mu+\delta\beta/\gamma$ and $\text{Var}(Y)=\delta\alpha^2/\gamma^3$, respectively. Here, $\delta$ is a scaling parameter, $\mu$ is a location parameter, $\beta$ controls the asymmetry, and $\alpha \pm \beta$ determines the heaviness of the tails. The density of the IG distribution with parameters $\delta$ and $\gamma$ is
\begin{equation}\label{eqn:fu}
f(u) =(2\pi)^{-1/2}\delta u^{-3/2}\exp\left\{\delta\gamma-\frac{1}{2}(\delta^2u^{-1}+\gamma^2u)\right\}.
\end{equation}
The expected value and variance of $U$ are $\ex[U]=\delta/\gamma$ and $\text{Var}[U]=\delta/\gamma^3$, respectively. Note that this is different from the parameterization of the IG distribution used by \cite{seshadri1993}, and can be obtained as a special case of generalized inverse Gaussian distribution \citep{chhikara1989}.
See \cite{barndorff97} and \cite{karlis09} for more details on the UNIG distribution.

\subsection{Parameter Estimation}

From \cite{karlis04}, the joint probability density is given by
$f(y,u)=f(u) f(y|u)$, where
$$ f(y|u) = (2\pi)^{-1/2} u^{-1/2}\exp {\left\{-\frac{1}{2u}(y-(\mu+\beta u))^2\right\}},$$
%$$f(u) =(2\pi)^{-1/2}\delta u^{-3/2}\exp\left\{\delta\gamma-\frac{1}{2}(\delta^2u^{-1}+\gamma^2u)\right\}.$$
and $f(u)$ is as defined in \eqref{eqn:fu}. Therefore,
$$f(y,u) \propto \delta \exp\left\{\delta\gamma-\beta\mu\right\}u^{-2}\exp\left\{\beta y +\mu \frac{y}{u}-\frac{1}{2}(\beta^2+\gamma^2)u-\frac{1}{2}(\mu^2+\delta^2)u^{-1}\right\}.$$ 
%Also, $U|y$ follows a generalized inverse Gaussian distribution, such that 
%\begin{eqnarray*}
%\ex[U_{ig}]&=&\frac{\delta_g\phi_{ig}(y_i)^{1/2}}{\alpha_g}\frac{K_0(\delta_g\phi_{ig}(y_i)^{1/2}\alpha_g)}{K_{-1}(\delta_g\phi_{ig}(y_i)^{1/2}\alpha_g)},\\
%\ex[U^{-1}_{ig}]&=&\frac{\alpha_g}{\delta_g\phi_{ig}(y_i)^{1/2}}\frac{K_{-2}(\delta_g\phi_{ig}(y_i)^{1/2}\alpha_g)}{K_{-1}(\delta_g\phi_{ig}(y_i)^{1/2}\alpha_g)}.
%\end{eqnarray*}
The likelihood of the complete UNIG data, i.e. the observed $\by$ and the latent $\bu$ such that $(\by,\bu)=(y_1,\ldots,y_n,u_1,\ldots,u_n)$, has the form
$$L(\btheta)=r(\btheta)^n\left[\prod_{i=1}^n h(y_i,u_i)\right]\exp\left\{{\sum_{j=1}^4\Phi_j(\btheta)t_j(\by,\bu)}\right\},$$
which is within the exponential family. Here, $r(\btheta)$ is the normalization constant that depends on $\btheta$, $h(y,u)$ is a continuous function of $(y,u)$, $\Phi_j(\btheta)$ is the $j$th natural parameter, and $t_j(\by,\bu)$ is the $j$th sufficient statistic with $\bu=(u_1,\ldots,u_n)$. For the UNIG model: $\Phi_1=\beta$, $\Phi_2=\mu$, $\Phi_3=\beta^2+\gamma^2$, and $\Phi_4=\mu^2+\delta^2$; and $t_1(\by,\bu)=\sum_{i=1}^ny_i$, $t_2(\by,\bu)=\sum_{i=1}^ny_iu_i$, $t_3(\by,\bu)=\frac{1}{2}\sum_{i=1}^nu_i$, and $t_4(\by,\bu)=\frac{1}{2}\sum_{i=1}^nu_i^{-1}$.
 If the conjugate prior distribution of $\btheta$ is of the form 
$$h(\btheta)\propto r(\btheta)^{a_0}\exp\left\{{\sum_{j=1}^4\Phi_j(\btheta)a_j}\right\},$$
then the posterior distribution is of the form
$$h(\btheta\mid\by)\propto r(\btheta)^{(a_0+n)}\exp \left\{\sum_{j=1}^4\Phi_j(\btheta)\left(a_j+t_j(\by,\bu)\right)\right\}.$$ 
Because ($\mu,\beta$) is independent of ($\delta,\gamma$), a bivariate prior normal distribution can be assigned to ($\mu,\beta$), a gamma prior distribution can be assigned to $\delta^2$, and a truncated normal prior conditional on $\delta$ can be assigned to $\gamma$ \citep{karlis04}. The values $a_j$ will be discussed shortly, in the mixture context. 

Now consider $n$ independent random variables $Y_1,\ldots,Y_n$ from a $G$-component mixture of UNIG distributions. The likelihood of the observed data $\by=(y_1,\ldots,y_n)$ from this mixture will have the form
\begin{equation*}%\label{eqn:ulike}
L(\btheta) = \prod_{i=1}^n \sum_{g=1}^G \pi_g f(y_i;\boldsymbol{\theta}_g),
\end{equation*}
where $\pi_g>0$ such that $\sum_{g=1}^G \pi_g =1$, $\btheta=(\btheta_1,\ldots,\btheta_g)$, and $f(\textbf{y};\boldsymbol{\theta}_g)$ is the $g$th component density given by a UNIG distribution with parameters $\boldsymbol{\theta}_g=(\alpha_g,\beta_g,\mu_g,\delta_g)$. Note that $\pi_1,\ldots,\pi_G$ are called mixing proportions.

Define a component indicator variable $Z_{ig}$ such that $z_{ig} =1$ if the observation $i$ belongs to component $g$ and $z_{ig} =0$ otherwise. %For notational convenience, we also define $\sz_i\in\{1,\ldots,G\}$ to denote the component to which observation $i$ belongs, e.g., suppose that $G=3$, then $\sz_i=2$ corresponds to $z_{i1}=0,z_{i2}=1,z_{i3}=0$. 
The complete-data, i.e., the observed $y_i$, the latent $u_{ig}$, and the missing $z_{ig}$, likelihood of a $G$-component mixture of UNIG distributions can be written
\begin{equation*}\begin{split}
L_{\text{c}}(\btheta)&=\prod_{g=1}^G\prod_{i=1}^n \left[\pi_g~f(y_i|u_{ig};\mu_g,\beta_g)~f(u_{ig};\delta_g,\gamma_g)\right]^{z_{ig}}\\
&=\prod_{g=1}^G\left[r(\btheta_g)^{{\sum_{i=1}^nz_{ig}}}\left(\prod_{i=1}^nh(y_i,u_{ig})\right)\exp\left\{{\sum_{j=1}^4\Phi_j(\btheta_g)t_j(\by,\bu_g)}\right\}\right],
\end{split}\end{equation*}
where $\bu_g=(u_{1g},\ldots,u_{ng})$.
If the conjugate prior distribution of $\btheta_g=(\pi_g,\mu_g,\beta_g,\delta_g,\gamma_g)$ is of the form 
$$h(\btheta_g)\propto r(\btheta_g)^{a_{g,0}^{(0)}}\exp\left\{{\sum_{j=1}^4\Phi_j(\btheta_g)a_{g,j}^{(0)}}\right\},$$
with hyperparameters taking initial values $(a_{g,0}^{(0)},a_{g,1}^{(0)},\ldots,a_{g,4}^{(0)})$, then the posterior distribution is of the form
$$h(\btheta_g\mid\by)\propto r(\btheta_g)^{(a_{g,0}^{(0)}+\sum_{i=1}^nz_{ig})}\exp \left\{\sum_{j=1}^4\Phi_j(\btheta_g)\left(a_{g,j}^{(0)}+t_j(\by,\bu_g)\right)\right\},$$
where %the hyperparameters of the posterior distributions of the mixtures of UNIG models are 
\begin{eqnarray*}
 a_{g,0}&=a_{g,0}^{(0)}+\sum_{i=1}^nz_{ig}, \qquad\qquad & \qquad
 a_{g,1}=a_{g,1}^{(0)}+\sum_{i=1}^nz_{ig}y_i,\\
 a_{g,2}&=a_{g,2}^{(0)}+\sum_{i=1}^nz_{ig}u^{-1}_{ig}y_i, \qquad & \qquad
 a_{g,3}=a_{g,3}^{(0)}+0.5\sum_{i=1}^nz_{ig}u_{ig},\\
 a_{g,4}&=a_{g,4}^{(0)}+0.5\sum_{i=1}^nz_{ig}u^{-1}_{ig}. &~\\
 \end{eqnarray*}

The approximating density in the variational Bayes framework is restricted to a factorized form for computational convenience, so that $q_{\btheta,\bw}(\boldsymbol{\theta},\mathbf{w})=q_{\btheta}(\boldsymbol{\theta})q_{\bw}(\mathbf{w})$.  For the mixture of UNIG distributions, the missing data are $\mathbf{w}=(\mathbf{z},\mathbf{u})$, where $\bz=(\bz_1,\ldots,\bz_n)$ with $\bz_i=(z_{i1},\ldots,z_{iG})$, and $\bu$ is defined similarly. Therefore, the approximating density is $q_{\btheta,\bw}(\boldsymbol{\theta},\mathbf{w})=q_{\btheta}(\boldsymbol{\theta})q_{\bz,\bu}(\mathbf{z},\mathbf{u})$. Upon choosing a conjugate prior, the appropriate hyperparameters for the approximating density $q_{\btheta}(\boldsymbol{\theta})$ for data from an exponential-family model can easily be obtained.

A Dirichlet prior with initial hyperparameters $(a_{1,0}^{(0)},\ldots,a_{G,0}^{(0)})$ is assigned to the mixing components $\boldsymbol{\pi}=(\pi_1,\ldots,\pi_g)$ and results in a Dirichlet posterior distribution with hyperparameters $(a_{1,0},\ldots,a_{G,0})$. A bivariate normal prior distribution is assigned to $(\mu_g,\beta_g)$ such that
\begin{equation*}
 \left(
\begin{array}{c}
\mu_g \\
\beta_g
\end{array} \right) \sim \text{N} \left[ \left(
\begin{array}{c}
\bar{\mu}_g^{(0)} \\
\bar{\beta}_g^{(0)}
\end{array} \right),  \left(\begin{array}{cc}
\sigma^{{(0)}~2}_{\mu_g}& \rho_g^{(0)}\sigma_{\mu_g}^{(0)}\sigma_{\beta_g}^{(0)}\\
\rho_g^{(0)}\sigma_{\mu_g}^{(0)}\sigma_{\beta_g}^{(0)},&\sigma^{{(0)}~2}_{\beta_g}
\end{array}\right) \right],
\end{equation*}
where 
\begin{eqnarray*}
 \rho_g^{(0)}&=-\frac{a_{g,0}^{(0)}}{2\sqrt{a_{g,3}^{(0)}a_{g,4}^{(0)}}},\qquad&\qquad
 \bar{\mu}_g^{(0)}=\frac{1}{2(1-\rho_g^{{(0)}~2})a_{g,4}^{(0)}}\left(a_{g,2}^{(0)}-\frac{a_{g,0}^{(0)}a_{g,1}^{(0)}}{2a_{g,3}^{(0)}}\right),\\
\sigma^{{(0)}~2}_{\mu_g}&=\frac{1}{2(1-\rho_g^{{(0)}~2})a_{g,4}^{(0)}} ,\qquad &\qquad
\bar{\beta}_g^{(0)}=\frac{1}{2(1-\rho_g^{{(0)}~2})a_{g,3}^{(0)}}\left(a_{g,1}^{(0)}-\frac{a_{g,0}^{(0)}a_{g,2}^{(0)}}{2a_{g,4}^{(0)}}\right),\\
 \sigma^{{(0)}~2}_{\beta_g}&=\frac{1}{2(1-\rho_g^{{(0)}~2})a_{g,3}^{(0)}}.&\\
 \end{eqnarray*}
The resulting posterior distribution for ($\mu_g,\beta_g$) is
\begin{equation*}
 \left(
\begin{array}{c}
\mu_g \\
\beta_g
\end{array} \right) \sim \text{N} \left[ \left(
\begin{array}{c}
\bar{\mu}_g \\
\bar{\beta}_g
\end{array} \right),  \left(\begin{array}{cc}
\sigma^2_{\mu_g}& \rho_g\sigma_{\mu_g}\sigma_{\beta_g}\\
\rho_g\sigma_{\mu_g}\sigma_{\beta_g},&\sigma^2_{\beta_g}
\end{array}\right) \right],
\end{equation*}
 where \begin{eqnarray*}
 \rho_g&=-\frac{a_{g,0}}{2\sqrt{a_{g,3}a_{g,4}}},\qquad&\qquad
 \bar{\mu}_g=\frac{1}{2(1-\rho_g^2)a_{g,4}}\left(a_{g,2}-\frac{a_{g,0}a_{g,1}}{2a_{g,3}}\right),\\
\sigma^2_{\mu_g}&=\frac{1}{2(1-\rho_g^2)a_{g,4}} ,\qquad &\qquad
\bar{\beta}_g=\frac{1}{2(1-\rho_g^2)a_{g,3}}\left(a_{g,1}-\frac{a_{g,0}a_{g,2}}{2a_{g,4}}\right),\\
 \sigma^2_{\beta_g}&=\frac{1}{2(1-\rho_g^2)a_{g,3}}.&\\
 \end{eqnarray*}

A gamma prior distribution is assigned to $\delta^2$ and a truncated normal prior distribution conditional on $\delta_g$ is assigned to $\gamma_g$, i.e.,
 \begin{equation*}
 \delta_g^2 \sim \mbox{Gamma} \left( \frac{a_{g,0}^{(0)}}{2}+1,a_{g,4}^{(0)}-\frac{a_{g,0}^{{(0)}~2}}{4a_{g,3}^{(0)}} \right)
 \end{equation*}
 and 
 \begin{equation*}
 \gamma_g\mid\delta_g \sim \text{N}\left( \frac{a_{g,0}^{(0)}\delta_g}{2a_{g,3}^{(0)}},\frac{1}{2a_{g,3}^{(0)}}\right )I(\gamma_g>0).
 \end{equation*}
 The resulting posterior distribution for ($\delta_g,\gamma_g$) is given by 
 \begin{equation*}
 \delta_g^2 \sim \mbox{Gamma} \left( \frac{a_{g,0}}{2}+1,a_{g,4}-\frac{a_{g,0}^2}{4a_{g,3}} \right)
 \end{equation*}
 and 
 \begin{equation*}
 \gamma_g\mid\delta_g \sim \text{N}\left( \frac{a_{g,0}\delta_g}{2a_{g,3}},\frac{1}{2a_{g,3}}\right )I(\gamma_g>0).
 \end{equation*}
 
For the variational approximation, $h(\btheta,\mathbf{w}\mid\by)$ is taken to have a factorized form, $q_{\btheta,\bw}(\boldsymbol{\theta},\mathbf{w})=q_{\btheta}(\boldsymbol{\theta})q_{\bw}(\mathbf{w})=q_{\btheta}(\boldsymbol{\theta})q_{\bz,\bu}(\mathbf{z},\mathbf{u})$. Following \cite{beal2003}, $q_{\bz_g,\bu_g}(\mathbf{z}_g=\mathbf{1},\mathbf{u}_g)$ for the conjugate-exponential models can be obtained as
$$q_{\bz_g,\bu_g}(\mathbf{z}_g=\mathbf{1},\mathbf{u}_g)=\prod_{i=1}^nq_{z_{ig},u_{ig}}(z_{ig}=1,u_{ig})$$ 
and
\begin{equation*}\begin{split}
q_{z_{ig},u_{ig}}(z_{ig}=1,u_{ig})&=\frac{1}{\mathcal{Z}_{z_{ig},u_{ig}}} \exp\left\{\int_\theta \log~p(y_i,z_{ig}=1,u_{ig}|\boldsymbol{\theta}_g)q_{\btheta}(\boldsymbol{\theta}_g) d\btheta_g\right\}\\
&=\frac{1}{\mathcal{Z}_{z_{ig},u_{ig}}} \exp\left\{\ex[\log~p(y_i,z_{ig}=1,u_{ig}|\boldsymbol{\theta}_g)]_{q_{\btheta}(\boldsymbol{\theta}_g)}\right\},
\end{split}\end{equation*}
where $\mathcal{Z}_{z_{ig},u_{ig}}$ is a constant.

The log of the mixture density given the parameters $\btheta_g$ is
\begin{equation*}\begin{split}
\log~p&(y_i,z_{ig}=1,u_{ig}|\boldsymbol{\theta}_g)=\log(\pi_g)-\log(2\pi)-2 \log (u_{ig})+\log (\delta_g)\\
&+\delta_g\gamma_g+\left(y_i-\mu_g\right)\beta_g-\frac{1}{2}\left[\left(\delta_g^2+(y_i-\mu_g)^2\right)u^{-1}+\left(\gamma_g^2+\beta_g^2\right)u\right].
\end{split}\end{equation*}
Setting $A_{ig}=\delta_g^2+(y_i-\mu_g)^2$, $B_g=\gamma_g^2+\beta_g^2$ and $C_{ig}=\delta_g\gamma_g+\left(y_i-\mu_g\right)\beta_g$, we can write
\begin{equation*}\begin{split}
\log~p(y_i,z_{ig}=1,u_{ig}|\boldsymbol{\theta})=\log(\pi_g)-\log(2\pi)-&2\log (u_{ig})+\log (\delta_g)+C_{ig}\\&-\frac{1}{2}\left[A_{ig}u_{ig}^{-1}+B_gu_{ig}\right].
\end{split}\end{equation*}
Hence, 
\begin{equation*}\begin{split}
\ex[\log~p(y_i,z_{ig}=1,u_{ig}|\boldsymbol{\theta})]=\ex[\log(\pi_g)]-\log(2\pi)&-2 \log (u_{ig})+\ex[\log (\delta_g)]+\ex[C_{ig}]\\
&-\frac{1}{2}\left[\ex[A_{ig}]u_{ig}^{-1}+\ex[B_g]u_{ig}\right].
\end{split}\end{equation*}
Therefore,
\begin{equation*}\begin{split}
& q_{z_{ig},u_{ig}}(z_{ig}=1,u_{ig})\propto\exp \left\{\ex[\log(\pi_g)]-\log(2\pi)-2 \log (u_{ig})+\ex[\log (\delta_g)]+\ex[C_{ig}]\right\}\\
 &\qquad\qquad\qquad\qquad\qquad+\exp\left\{-\frac{1}{2}\left[\ex[A_{ig}]u_{ig}^{-1}+\ex[B_g]u_{ig}\right]\right\}\\
 &=(2\pi)^{-1}\exp\left\{\ex[\log(\pi_g)]+\ex[\log (\delta_g)]+\ex[C_{ig}]\right\}u_{ig}^{-2}\exp\left\{-\frac{1}{2}\left[\ex[A_{ig}]u_{ig}^{-1}+\ex[B_g]u_{ig}\right]\right\}\\
 &=(2\pi)^{-1}\exp\left\{\ex[\log(\pi_g)]+\frac{1}{2}\ex[\log (\delta_g^2)]+\ex[C_{ig}]\right\}\text{GIG}\left(u_{ig}~\bigg|~-1,\sqrt{\ex[A_{ig}]},\sqrt{\ex[B_g]}\right).
\end{split}\end{equation*}
Here, GIG$(\cdot)$ is the probability density function of generalized inverse Gaussian distribution \citep{jorgensen1982} and
\begin{equation*}\begin{split}
& \ex[\log(\pi_g)]=\Psi(a_{g,0})-\Psi(n) ,\\
& \ex[\log (\delta_g^2)]=\Psi\left(({a_{g,0}}/{2})+1\right)-\log\left(a_{g,4}-({a_{g,0}^2}/{4a_{g,3}})\right),\\
& \ex[A_{ig}]=\ex[\delta_g^2]+\ex[(y_i-\mu_g)^2]
=\frac{({a_{g,0}}/{2})+1}{a_{g,4}-({a_{g,0}^2}/{4a_{g,3}})}+y_i^2-2y_i\ex[\mu_g]+\ex[\mu_g^2],\\
& \ex[B_g]=\ex[\gamma_g^2]+\ex[\beta_g^2]
=(\ex[\gamma_g^2])^2+\text{Var}(\gamma_g)+ (\ex[\beta_g^2])^2+\text{Var}(\beta_g),\\
& \ex[C_{ig}]=\ex[\delta_g\gamma_g]+\ex[\left(y_i-\mu_g\right)\beta_g]
=\ex[\delta_g\gamma_g]+y_i\ex[\beta_g]-\left(\ex[\mu_g]\ex[\beta_g]+\text{Cov}(\mu_g,\beta_g)\right),\\
\end{split}\end{equation*} where $\Psi(\cdot)$ is the digamma function.
 
 The approximating density $q_{z_{ig}}(z_{ig}=1)$ %, for $i = 1,\ldots,n$ and $g= 1,\ldots,G$, 
 is
\begin{equation*}\begin{split}
   q_{z_{ig}}(z_{ig}=1)&=\int_{u_{ig}} q_{z_{ig},u_{ig}}(z_{ig}=1,u_{ig}) du_{ig}\\
   &\propto\int_{u_{ig}}(2\pi)^{-1}\exp\left\{\ex[\log(\pi_g)]+\frac{1}{2}\ex[\log (\delta_g^2)]+\ex[C_{ig}]\right\}\\
   &\qquad\qquad\qquad\times\text{GIG}\left(u_{ig}~\bigg|~-1,\sqrt{\ex[A_{ig}]},\sqrt{\ex[B_g]}\right)du_{ig}\\
   &=(2\pi)^{-1}\exp\left\{\ex[\log(\pi_g)]+\frac{1}{2}\ex[\log (\delta_g^2)]+\ex[C_{ig}]\right\}\\
   &\qquad\qquad\qquad\times 2\left(\frac{\ex[A_{ig}]}{\ex[B_g]}\right)^{-1/2}K_{-1}\left(\sqrt{\ex[A_{ig}]\ex[B_g]}\right).
\end{split}\end{equation*} %where $K_{1}(y)$ is the modified Bessel function of the third kind of order 1 evaluated at $y$.
Using the approximating density $q_{z_{ig}}(z_{ig}=1)$, the probability that observation $i$ belongs to component $g$ is
 $$\hat{z}_{ig}=\frac{q_{z_{ig}}(z_{ig}=1)}{\sum_{g=1}^Gq_{z_{ig}}(z_{ig}=1)}.$$
The approximating density $q_{u_{ig}}(u_{ig}\mid z_{ig}=1)$ is
\begin{eqnarray*}
q_{u_{ig}}(u_{ig}\mid z_{ig}=1) &\propto& (2\pi)^{-1}\exp\left\{\ex[\log(\pi_g)]+\frac{1}{2}\ex[\log (\delta_g^2)]+\ex[C_{ig}]\right\}\\
&\times& 2\left(\frac{\ex[A_{ig}]}{\ex[B_g]}\right)^{-1/2}K_{-1}\left(\sqrt{\ex[A_{ig}]\ex[B_g]}\right),\\
   \end{eqnarray*}
and so $U_{ig}\mid(z_{ig}=1)\backsim\text{GIG}(-1,\sqrt{\ex[A_{ig}]},\sqrt{\ex[B_g]})$. Therefore,
\begin{eqnarray*}
\ex[U_{ig}\mid z_{ig}=1]_{q_{u_{ig}}(u_{ig}\mid z_{ig}=1)}&=&\left(\frac{\ex[A_{ig}]}{\ex[B_g]}\right)^{1/2}\frac{K_0(\sqrt{\ex[A_{ig}]\ex[B_g]})}{K_{-1}(\sqrt{\ex[A_{ig}]\ex[B_g]})},\\
\ex[U^{-1}_{ig}\mid z_{ig}=1]_{q_{u_{ig}}(u_{ig}\mid z_{ig}=1)}&=&\left(\frac{\ex[A_{ig}]}{\ex[B_g]}\right)^{-1/2}\frac{K_{-2}(\sqrt{\ex[A_{ig}]\ex[B_g]})}{K_{-1}(\sqrt{\ex[A_{ig}]\ex[B_g]})}.
\end{eqnarray*}

The variational Bayes algorithm proceeds in the following manner: 
 \begin{itemize}
\item For the observed data $\by=(y_1,y_2,\ldots,y_n)$, the algorithm is initialized with more components than expected, say $G$. The $\hat{z}_{ig}$ can be initialized by either randomly assigning the observations to one of the $G$ components or by using the results from another clustering method (e.g., $k$-means clustering). 
\item Using the initialized values of $\hat{z}_{ig}$, the parameters from the $g$th component are initialized as follows:
\begin{itemize}
\item The component's sample mean is used to initialize the parameter $\mu_g$, 
\item $\beta_g$ is set to 0, and 
\item $\gamma$ and $\delta$ are set to 1.
\end{itemize} 
\item Using these values of the parameters, the expected values of  $U^{-1}_{ig}$ and $U_{ig}$ are initialized.
\item The hyperparameters of the prior distributions are initialized to give a flat distribution over the possible values of the parameters. In our case, we chose $a_{g,j}^{(0)}=10^{-8}$, for $j=0,\ldots,4$; see Section~\ref{sec:sim3} for a simulation study that investigates sensitivity to our choice of $10^{-8}$.
\begin{enumerate}
\item Using the $\hat{z}_{ig}$ and the expected values of $U^{-1}_{ig}$ and $U_{ig}$, the hyperparameters of the approximating density $q_{\btheta}(\btheta)$ are updated. Using these updated hyperparameters, %of $q_{\btheta}(\btheta)$, 
the expected values $\ex[\log r(\btheta)]$ and $\ex[\phi_j(\btheta)]$ %, $j =1,\ldots,4$, 
are updated.\label{step1}
\item Using these updated $\ex[\log r(\btheta)]$ and $\ex[\phi_j(\btheta)]$, %$j =1,\ldots,4$, 
the $\hat{z}_{ig}$, $\ex[U^{-1}_{ig}|z_{ig}=1]$, and $\ex[U_{ig}|z_{ig}=1]$ are updated.\label{step2}
\item Components with too few observations are eliminated. Specifically, for each component $g'$ we do the following. If the estimated number of observations in component $g'$, i.e., $\sum_{i=1}^n\hat{z}_{ig'}$, is sufficiently small (less than one in our case), then component $g'$ is eliminated.\label{step3}
\end{enumerate}
Steps \ref{step1},  \ref{step2}, and  \ref{step3} are repeated until convergence.
\end{itemize}

Once convergence is achieved, the observations are assigned to clusters using maximum \textit{a posteriori} probability (MAP), such that $\text{MAP}(\hat{z}_{ig})=1$ if $\max_g(\hat{z}_{ig})$ occurs in component $g$ and $\text{MAP}(\hat{z}_{ig}) = 0$ otherwise. If the true class is known, as in our analysis, the performance of the algorithm can be assessed using the adjusted Rand index \citep[ARI;][]{hubert85}. The ARI is based on the pairwise agreement between the predicted and true classifications after adjusting for agreement by chance: a value of `1' indicates a perfect classification and a value of `0' would be expected under random classification.

\subsection{Simulated Data}
\subsubsection{Simulation Study 1}
\begin{figure}[htbp]
\includegraphics[width=0.975\textwidth]{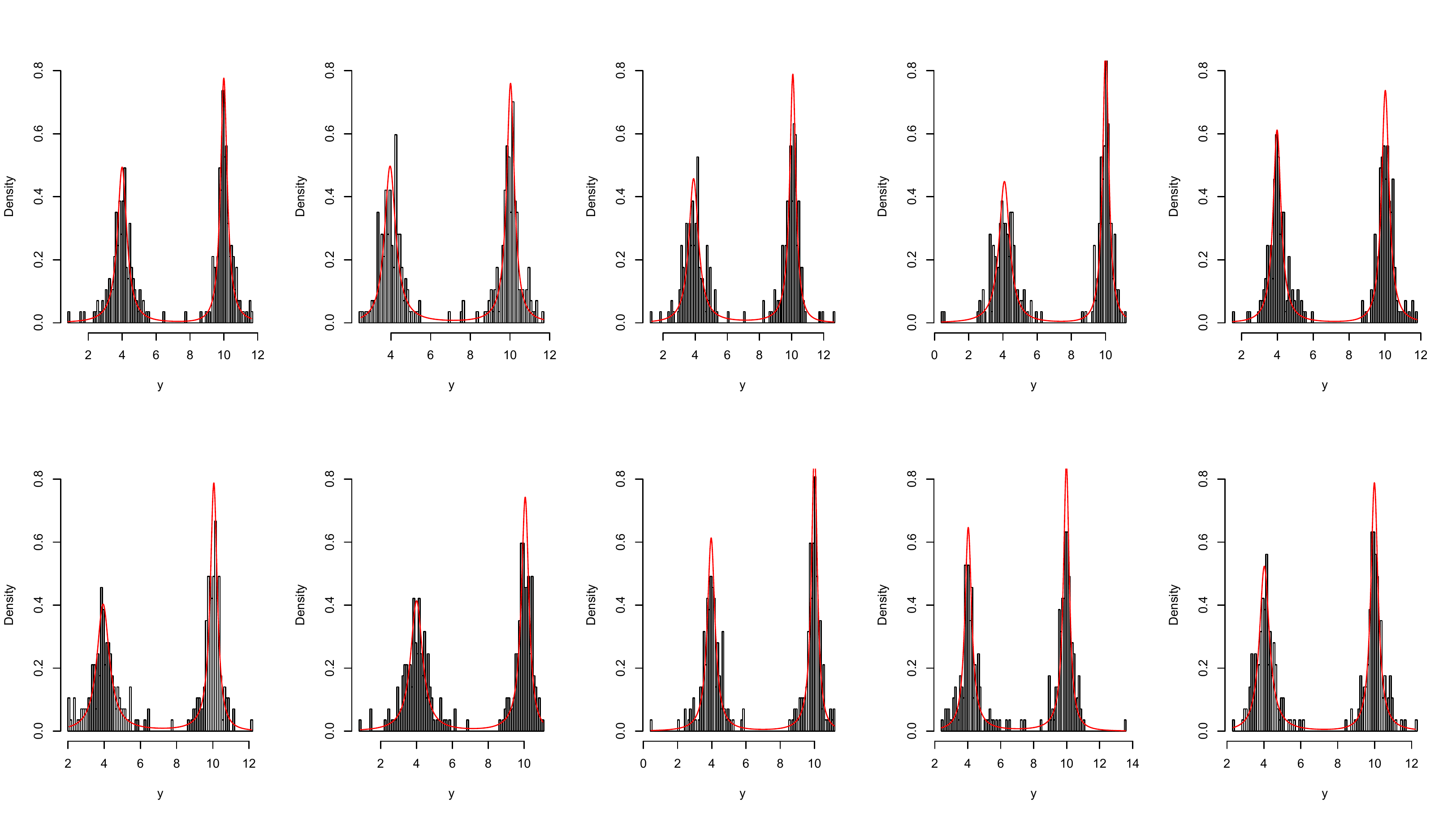}
\caption{Histograms with fitted densities for ten data sets from Simulation~1.}
\label{fig1}
\end{figure}

We simulated one-hundred data sets from a UNIG mixture with two components ($n_1=150$ and $n_2=150$). We chose the parameters %(Table~\ref{tab1}) 
so that the components are well separated. We ran our variational Bayes algorithm starting off with $G=10$ components. In all one-hundred cases, our approach gave a two-component model and classification was excellent ($\text{mean ARI} = 0.99$ with $\text{std.\ dev.} = 0.01$). %The results (Table~\ref{tab1}) indicate that the average estimates of the component location parameters $(\mu_1,\mu_2)$ are very close to the true values. Interpreting how good the other parameter estimates are is difficult; however, 
Looking at the predicted density for ten of the simulated data sets (Figure~\ref{fig1}), it is clear that the fitted densities are capturing the data very well.
%\begin{table}[htbp]
%\begin{center}
%\caption{True and (average) estimated values of the parameters from the univariate simulation.}
%\begin{tabular*}{1.0\textwidth}{@{\extracolsep{\fill}}llrrrr}
%\hline
%&& $\alpha$ &$\beta$ & $\delta$ &$\mu$\\
%\hline
%\multirow{2}{*}{Group 1} & True&2&0.75&0.5&5\\
%& Estimated&0.73&0.45&0.24&5.06\\
%\hline
%\multirow{2}{*}{Group 2} &True&1.5&0&0.24&10\\
% &Estimated&0.62&-0.17&0.16&10.01\\
%\hline
%\end{tabular*}
%\label{tab1}
%\end{center}
%\end{table}

\subsubsection{Simulation Study 2}
\begin{figure}[htbp]
\includegraphics[width=0.975\textwidth]{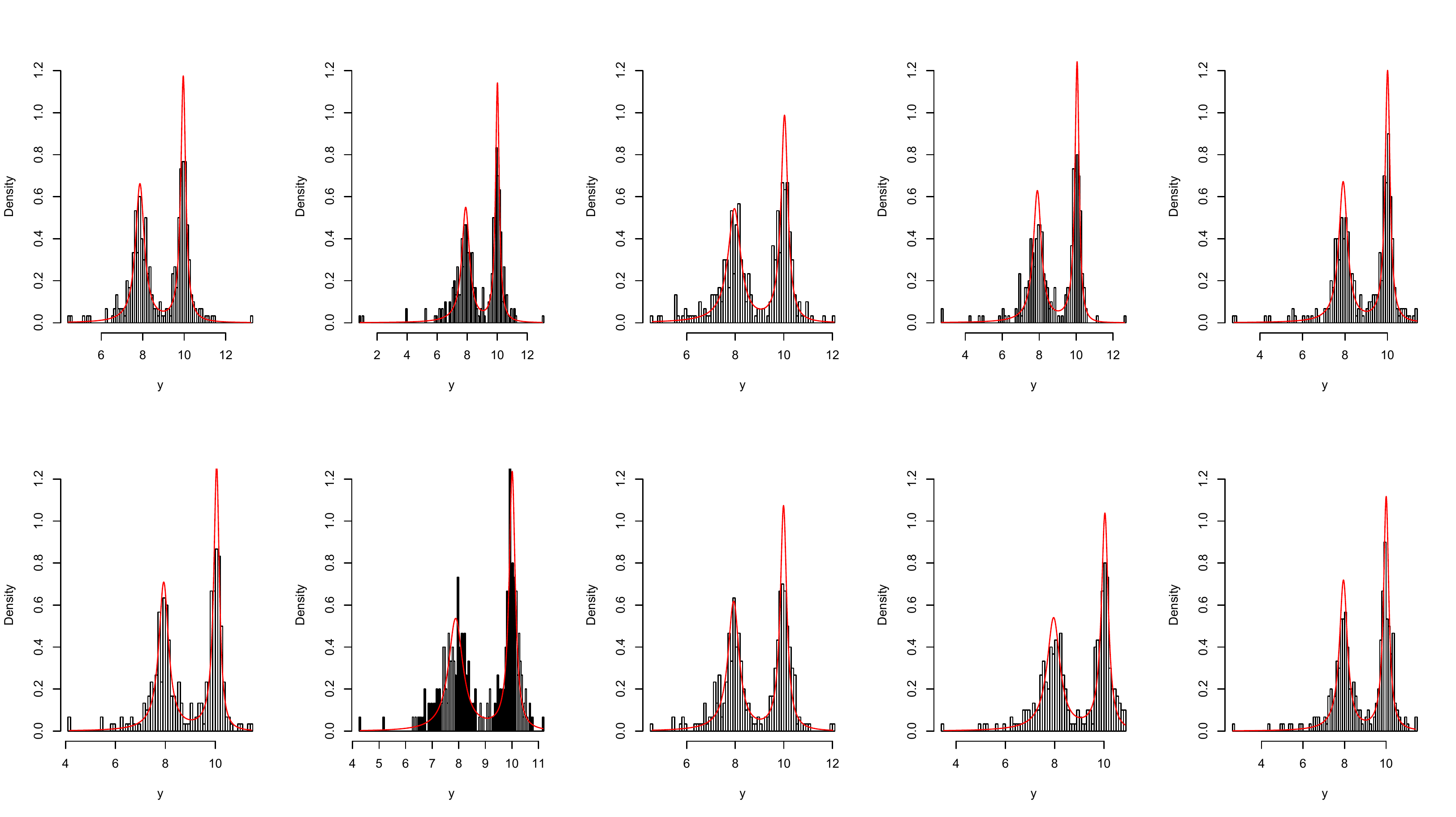}
\caption{Histograms with fitted densities for ten data sets from Simulation~2.}
\label{figsim2}
\end{figure}

We simulated another one-hundred data sets from a UNIG mixture with two components ($n_1=150$ and $n_2=155$). In this case, the components were not as well separated. We again ran our variational Bayes algorithm starting off with $G=10$ components. Out of the one-hundred data sets, a two-component model was selected on 92 occasions and the mean ARI over all one-hundred data sets is $0.92$ (with standard deviation $0.03$). Figure~\ref{figsim2} shows the fitted densities for ten of the simulated data sets; again, the fitted densities are capturing the data very well.

\subsubsection{Simulation Study 3}\label{sec:sim3}
\begin{figure}[htbp]
\vspace{-0.2in}
\includegraphics[width=0.975\textwidth]{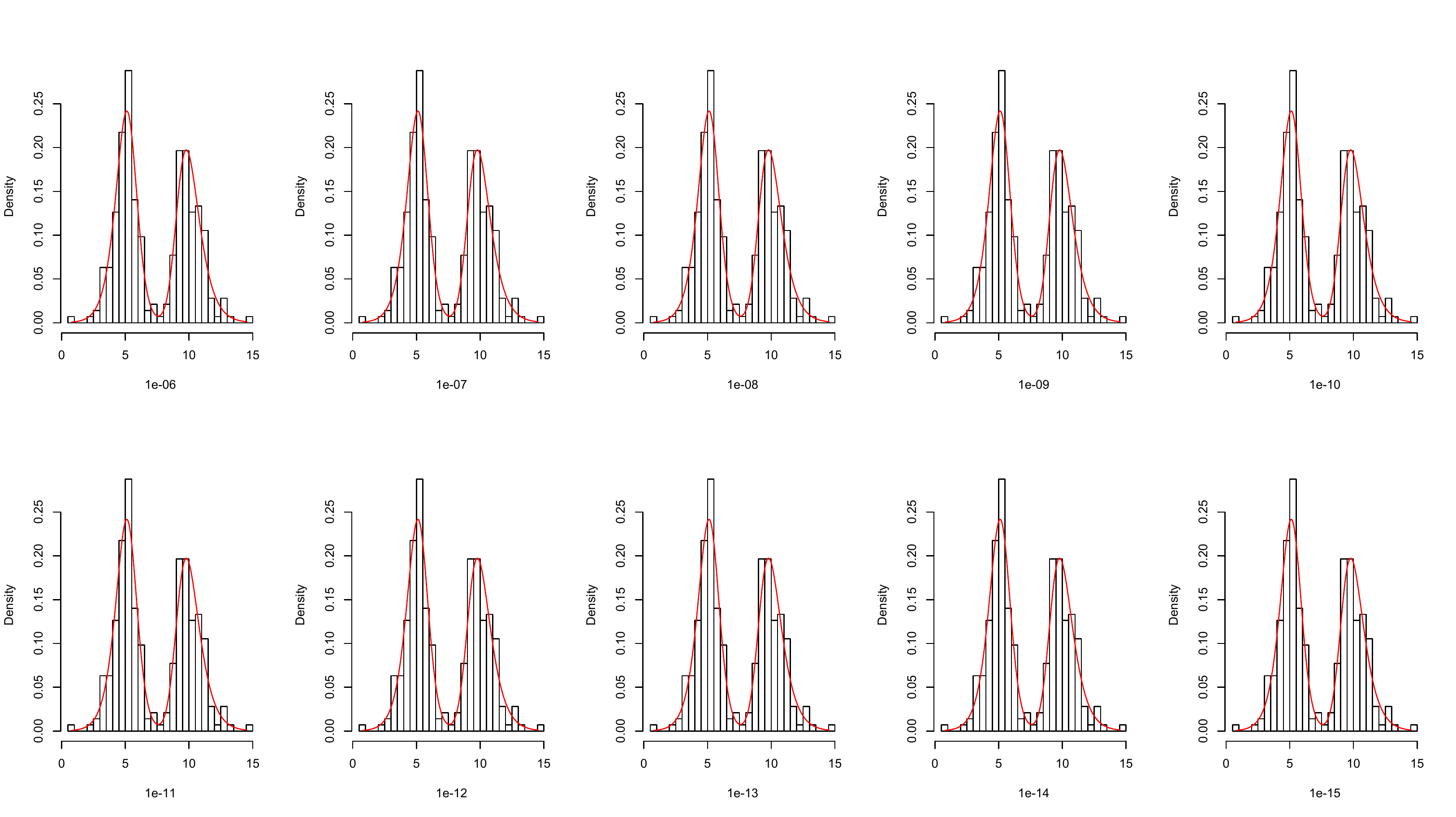}
\caption{Histograms with fitted densities for the ten data sets from Simulation~3, where the label of each $x$-axis reflects the initial values for the hyperparameters.}
\label{figsim5}
\end{figure}

Recall that we initialize hyperparameters for $\btheta_g=(\pi_g,\mu_g,\beta_g,\delta_g,\gamma_g)$ so that the prior distribution of $\btheta_g$ is relatively flat. To evaluate the effect of the choice of initial values for these hyperparameters, we ran our algorithm on simulated data using 10 different initializations for these hyperparameters. Specifically, we used initial values $$a_{0g} =a_{1g}=a_{2g} =a_{3g}=a_{4g} \in\{10^{-6}, 10^{-7}, \ldots, 10^{-15}\},$$ for $g=1,\ldots,G$. For each of the ten runs, the data and the initial $\hat{z}_{ig}$ were the same so that only the initial values of the hyperparameters differed. The classification results obtained from all ten different initial values for the hyperparameters are identical ($\text{ARI}=0.99$), and the fitted densities for all ten runs are virtually identical (Figure~\ref{figsim5}). 

\subsection{Enzyme Data Set}
\begin{figure}[htbp]
\vspace{-0.2in}
\includegraphics[width=0.9\textwidth]{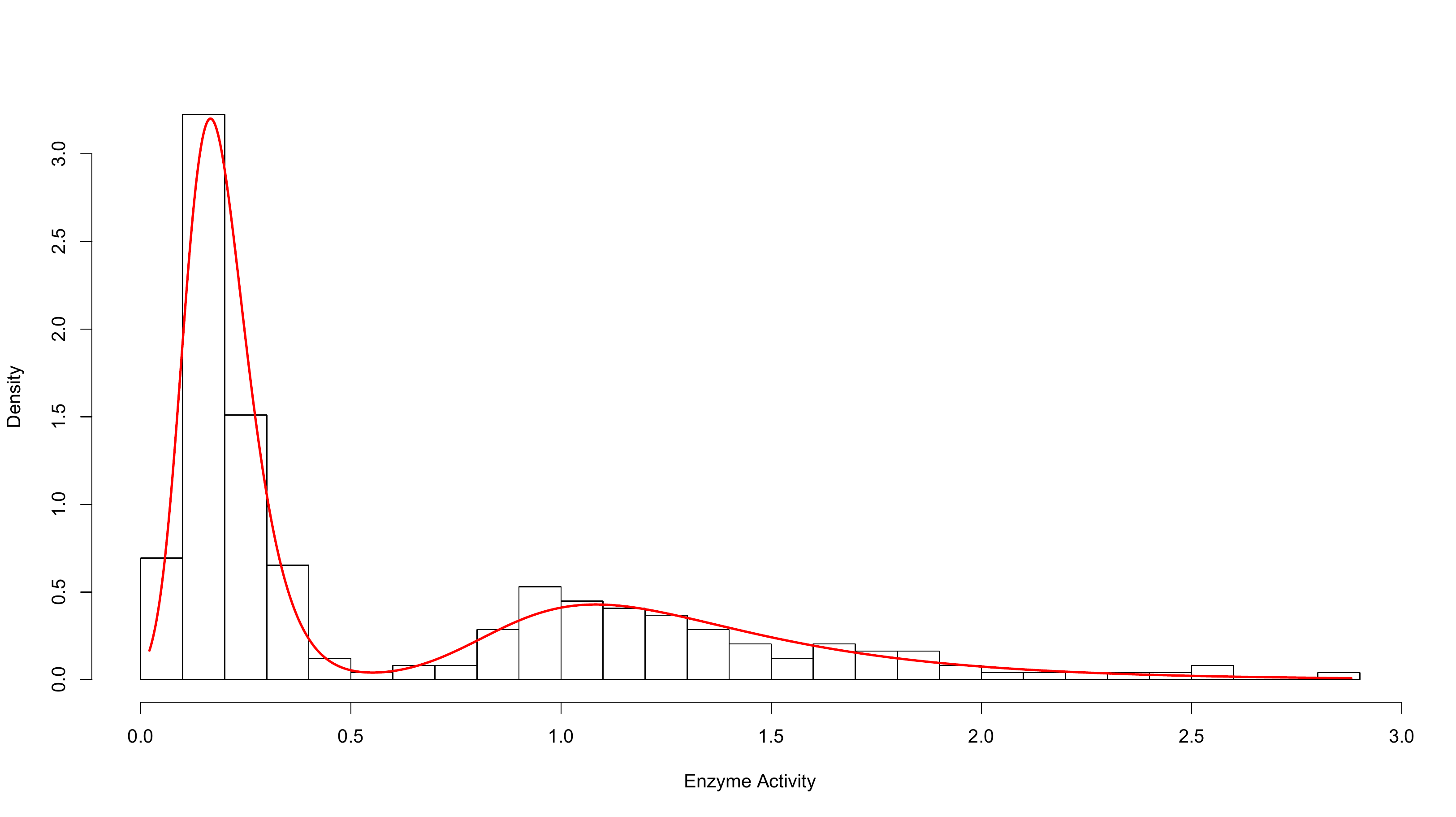}
\caption{Histogram with fitted density for the enzyme data.}
\label{fig2}
\end{figure}

We considered the enzyme data set, which is a benchmark data set for a mixture of univariate distributions with a skewed component \citep{bechtel93,karlis09}. The data consist of measurements of the activity of an enzyme in the blood of $245$ individuals. These data were used by \cite{karlis09} to illustrate fitting of the UNIG models within an EM algorithm framework. Their EM algorithm, in conjunction with a model selection criterion, resulted in the selection of a two-component UNIG model. We used our variational Bayes approach to fit the UNIG models, initializing at $G=5$ components. Akin to \cite{karlis09}, we obtained a two-component model that clearly gives a good fit to the data (Figure~\ref{fig2}).

%%% MNIG begins here

\section{Mixture of Multivariate Normal Inverse Gaussian Distributions}
\label{sec:3}
\subsection{The Model}
%Let us consider $n$ independent random variables $\bY_1,\ldots,\bY_n$ such that $\bY\in\mathcal{R}^d$. 
A mean-variance mixture of a $d$-dimensional multivariate normal distribution with the inverse Gaussian distribution, i.e.,
\begin{equation*}
\mathbf{Y} \mid w \sim \text{N}(\bm{\mu}+w\boldsymbol{\beta}\bm{\Delta}, w\bm{\Delta}), \qquad
%\end{equation*}
%\begin{equation*}
W \sim \text{IG}(\delta,\gamma)
\end{equation*}
 results in a MNIG distribution with density
\begin{equation*}
f(\mathbf{y} ; \theta)=\frac{\delta}{2^\frac{d-1}{2}}\exp\left\{\delta\gamma+(\mathbf{y}-\boldsymbol{\mu})\boldsymbol{\beta}'\right\} \left[\frac{\alpha}{\pi q(\mathbf{y} )}\right]^\frac{d+1}{2}K_{\frac{d+1}{2}}(\alpha q(\mathbf{y} )),
\end{equation*}
where $\alpha^2=\gamma^2+\Beta\bm{\Delta}\Beta'$, $q(\mathbf{y} )^2=\delta^2+(\mathbf{y} -\boldsymbol{\mu})\bm{\Delta}^{-1}(\mathbf{y} -\boldsymbol{\mu})'$, and $K_r(\mathbf{y})$ is the modified Bessel function of the third kind of order $r$ evaluated at $\mathbf{y}$.
Similar to the univariate case, the parameters contribute to the different shapes the MNIG can have. Here, $\bm{\Delta}$ is a $d \times d$ symmetric positive definite matrix that relates to the covariance matrix via
$$\text{Cov}(\bY)=\frac{\delta}{\gamma^3}(\gamma^2\bm{\Delta}+\bm{\Delta}\Beta'\Beta\bm{\Delta}),$$ and the restriction $|\bm{\Delta}| = 1$ is needed to ensure identifiability.
Conjugate priors are unavailable for $\bm\Delta$ with this restriction ($|\bm{\Delta}| = 1$). An alternative re-parameterization, as discussed in \cite{karlis09}, arises from
\begin{equation*}
\mathbf{Y} \mid u \sim \text{N}(\tilde{\Mu}+u\tilde{\Beta}, u\tilde{\bm{\Sigma}}), \qquad U \sim \text{IG}(1,\tilde{\gamma}),
\end{equation*}
 where $\tilde{\Mu}=\Mu$, $\tilde{\gamma} = \gamma \delta$, $\tilde{\bm{\Sigma}}=\delta^2\bm{\Delta}$, and $\tilde{\Beta}=\Beta \tilde{\bm{\Sigma}}$. Here, $\tilde{\bm{\Sigma}}$ is not restricted and conjugate priors exist for all of the model parameters.
 
\subsection{Parameter Estimation}
The joint probability density is $f(\mathbf{y} ,u)=f(u) f(\mathbf{y} |u),$ 
where
$$ f(\mathbf{y} |u) = (2\pi)^{-1/2} u^{-d/2}|\tilde{\bm{\Sigma}}|^{-d/2}\exp\left\{\frac{-1}{2u}(\mathbf{y} -\tilde{\Mu}-\tilde{\Beta} u)'\tilde{\bm{\Sigma}}^{-1}(\mathbf{y} -\tilde{\Mu}-\tilde{\Beta} u)\right\}$$ and
 $$f(u) =(2\pi)^{-1/2}u^{-3/2}\exp\left\{\tilde{\gamma}-\frac{1}{2}(2u^{-1}+\tilde{\gamma}^2u)\right\}.$$
Therefore,
\begin{eqnarray*}f(\mathbf{y} ,u) &\propto& u^{-\frac{d+3}{2}}|\tilde{\bm{\Sigma}}|^{-d/2}\exp\left\{-\frac{1}{2u}(\mathbf{y} -\tilde{\Mu}-\tilde{\Beta} u)'\tilde{\bm{\Sigma}}^{-1}(\mathbf{y} -\tilde{\Mu}+\tilde{\Beta} u)\right\}\\ &\times&\exp\left\{-\frac{1}{2}(2u^{-1}+\tilde{\gamma}^2u-2\tilde{\gamma})\right\}.
\end{eqnarray*}

If ${\boldsymbol{\theta}}=(\tilde{\Mu},\tilde{\bm\Sigma},\tilde{\Beta},\tilde{\gamma})$, then the likelihood of the complete MNIG data, i.e. the observed $\by$ and the latent $\bu$ such that $(\by,\bu)=(\by_1,\ldots,\by_n,u_1,\ldots,u_n)$, has the form
$$L(\btheta)=r(\boldsymbol{\theta})^{n}\left(\prod_{i=1}^nh(\mathbf{y}_i ,u_i)\right)\exp\left\{{\sum_{j=1}^4\Phi_j(\boldsymbol{\theta})t_j(\mathbf{y} ,\bu)}\right\},$$
which is within the exponential family. Note that, as before, $\bu=(u_1,\ldots,u_n)$. Here, 
$r(\btheta)$ is the normalization constant that depends on $\btheta$, $h(\by,u)$ is a continuous function of $(\by,u)$, $\Phi_j(\btheta)$ is the $j$th natural parameter, and $t_j(\by,\bu)$ is the $j$th sufficient statistic.

%If the conjugate prior distribution of $\boldsymbol{\theta}$ is of the form 
%$$p(\boldsymbol{\theta})\propto g(\boldsymbol{\theta})^{a_0}\exp\left\{{\sum_{j=1}^4\Phi_j(\boldsymbol{\theta})a_j}\right\},$$
%then the posterior distribution is of the form
%$$p(\boldsymbol{\theta}|\mathbf{y},u)\propto g(\boldsymbol{\theta})^{(a_0+n)}\exp \left\{{\sum_{j=1}^4\Phi_j(\boldsymbol{\theta})\left(a_j+t_j(\mathbf{y} ,u)}\right)\right\},$$
% which is the approximating density for the variational Bayes algorithm. 
 
Now, consider $n$ independent observations $\by_1,\ldots,\by_n$ from a $G$-component mixture of MNIG distributions. The likelihood is given by
$$L(\btheta) = \prod_{i=1}^n \sum_{g=1}^G \pi_g f(\mathbf{y};\tilde{\bm{\Sigma}}_g,\tilde{\Beta}_g,\tilde{\Mu}_g,\tilde{\gamma} _g),$$
where $f(\mathbf{y};\cdot)$ is the density of the MNIG distribution and the $\pi_g>0$, such that $\sum_{i=1}^G \pi_g =1$, are the mixing proportions. In this case, ($\tilde{\Mu}_g,\tilde{\bm\beta}_g$) is independent of $\tilde{\gamma}_g$. 

The complete-data likelihood for a $G$-component mixture of MNIG distributions can be written
\begin{eqnarray*}
L(\btheta)&=&\prod_{g=1}^G\prod_{i=1}^n \left[\pi_g~f(\by_i|u_{ig};\tilde{\Mu}_g,\tilde{\bbeta}_g,\tilde{\bm{\Sigma}}_g)~f(u_{ig};1,\tilde{\gamma}_g)\right]^{z_{ig}}\\
&=&\prod_{g=1}^G\left[r(\btheta_g)^{(\sum_{i=1}^n{z_{ig}})}\left(\prod_{i=1}^nh(\by_i,u_{ig})\right)\exp\left\{{\sum_{j=1}^4\Phi_j(\btheta_g)t_j(\by,\bu_{g})}\right\}\right],
\end{eqnarray*}
where $\bu_g=(u_{1g},\ldots,u_{ng})$.
If the conjugate prior distribution of $\btheta_g=(\pi_g,\tilde{\Mu}_g,\tilde{\Sigma}_g,\tilde{\Beta}_g,\tilde{\gamma}_g)$ is of the form 
$$h(\btheta_g)\propto r(\btheta_g)^{a_{g,0}^{(0)}}\exp\left\{{\sum_{j=1}^4\Phi_j(\btheta_g)a_{g,j}^{(0)}}\right\},$$
with initial hyperparameters $\{a_{g,0}^{(0)},\ba_{g,1}^{(0)},\mathbf{a}_{g,2}^{(0)},a_{g,3}^{(0)},a_{g,4}^{(0)}\}$, then the posterior distribution is of the form
$$h(\btheta_g\mid\by)\propto r(\btheta_g)^{(a_{g,0}^{(0)}+\sum_{i=1}^nz_{ig})}\exp \left\{\sum_{j=1}^4\Phi_j(\btheta_g)\left(\ba_{g,j}^{(0)}+t_j(\by,\bu_{g})\right)\right\},$$
where the hyperparameters of the posterior distributions of the mixtures of MNIG models are 
\begin{eqnarray*}
 a_{g,0}&=a_{g,0}^{(0)}+\sum_{i=1}^nz_{ig}, \qquad\qquad & \qquad
\ba_{g,1}=\ba_{g,1}^{(0)}+\sum_{i=1}^nz_{ig}\by_i,\\
 \ba_{g,2}&=\ba_{g,2}^{(0)}+\sum_{i=1}^nz_{ig}u^{-1}_{ig}\by_i, \qquad & \qquad
 a_{g,3}=a_{g,3}^{(0)}+\sum_{i=1}^nz_{ig}u_{ig},\\
 a_{g,4}&=a_{g,4}^{(0)}+\sum_{i=1}^nz_{ig}u^{-1}_{ig}. \qquad&\\
 \end{eqnarray*}
 Note that $\ba_{g,j}^{(0)}$ and $\ba_{g,j}$ for $j=1,2$ are vectors, and $a_{g,j}^{(0)}$ and $a_{g,j}$ for $j\in \{0,3,4\}$ are scalars. When referring to a value $j\in \{0,1,2,3,4\}$, we write $a_{g,j}^{(0)}$ and $a_{g,j}$.

A Dirichlet prior with initial hyperparameters $(a_{1,0}^{(0)},\ldots,a_{G,0}^{(0)})$ is assigned to the mixing proportions and results in a Dirichlet posterior distribution with hyperparameters $(a_{1,0},\ldots,a_{G,0})$.
 
A Wishart prior was assigned to the precision matrix of the $g$th component, i.e., $\tilde{\bm{\Sigma}}^{-1}_g\backsim\text{Wishart }(a_{g,0},\mathbf{V}_g^{(0)})$, resulting in a Wishart posterior $\tilde{\bm{\Sigma}}^{-1}_g\backsim\text{Wishart }(a_{g,0},\mathbf{V}_g')$ with 
\begin{eqnarray*}
\mathbf{V}_g&=&\mathbf{V}^{(0)}_g+\sum_{i=1}^n\hat{z}_{ig}u^{-1}_{ig}\mathbf{y}'\mathbf{y}-  \mathbf{a}_{g,2}'\tilde{\Mu}-\tilde{\Mu}' \mathbf{a}_{g,2}+ a_{g,4}\tilde{\Mu}'\tilde{\Mu}-\tilde{\Beta}' \mathbf{a}_{g,1}+a_{g,0}\tilde{\Beta}'\tilde{\Mu}- \mathbf{a}_{g,1}'\tilde{\Beta}\\&+&a_{g,0}\tilde{\Mu}'\tilde{\Beta}+a_{g,3}\tilde{\Beta}'\tilde{\Beta}.
\end{eqnarray*}

A correlated multivariate Gaussian prior distirbution conditional on the precision matrix was assigned to ($\tilde{\Mu}_g,\tilde{\Beta}_g$) such that
\begin{equation*}
 \left(
\begin{array}{c}
\tilde{\Mu}_g \\
\tilde{\Beta}_g
\end{array} \right)\Bigg|\tilde{\bm{\Sigma}}^{-1} \sim \text{N} \left[ \left(
\begin{array}{c}
\bar{\Mu}_g \\
\bar{\Beta}_g
\end{array} \right),  \left(\begin{array}{cc}
\tilde{\bm{\Sigma}}^{-1}_{\mu_g}& \tilde{\bm{\Sigma}}^{-1}_{\mu_g\beta_g}\\
 \tilde{\bm{\Sigma}}^{-1}_{\mu_g\beta_g},&\tilde{\bm{\Sigma}}^{-1}_{\beta_g}
\end{array}\right)^{-1} \right],
\end{equation*}
where  
\begin{eqnarray*}
\tilde{\bm{\Sigma}}^{-1}_{\mu_g}&=a_{g,4}^{(0)}\tilde{\bm{\Sigma}}_g^{-1},\qquad&\qquad
 \bar{\Mu}_g=\frac{a_{g,3}^{(0)}}{(a_{g,3}^{(0)}a_{g,4}^{(0)}-a_{g,0}^{{(0)}~2})}\left(\mathbf{a}^{(0)}_{g,2}-\frac{\mathbf{a}_{g,1}^{(0)}a_{g,0}^{(0)}}{a_{g,3}^{(0)}}\right),\\
\tilde{\bm{\Sigma}}^{-1}_{\beta_g}&=a_{g,3} ^{(0)}\tilde{\bm{\Sigma}}_g^{-1},\qquad&\qquad
 \bar{\Beta}_g=\frac{a_{g,4}^{(0)}}{(a_{g,3}^{(0)}a_{g,4}^{(0)}-a_{g,0}^{^{(0)}~2})}\left(\mathbf{a}_{g,1}^{(0)}-\frac{\mathbf{a}_{g,2}^{(0)}a_{g,0}^{(0)}}{a_{g,4}^{(0)}}\right),\\
 \tilde{\bm{\Sigma}}^{-1}_{\mu_g\beta_g}&=a_{g,0}^{(0)} \tilde{\bm{\Sigma}}_g^{-1}.&\\
 \end{eqnarray*}
The resulting posterior distribution for ($\tilde{\bm\mu}_g,\tilde{\bm\beta}_g$) is given by
\begin{equation*}
 \left(
\begin{array}{c}
\tilde{\Mu}_g\\
\tilde{\Beta}_g
\end{array} \right)\Bigg|\tilde{\bm{\Sigma}}^{-1} \sim \text{N} \left[ \left(
\begin{array}{c}
\bar{\Mu}_g \\
\bar{\Beta}_g
\end{array} \right),  \left(\begin{array}{cc}
\tilde{\bm{\Sigma}}^{-1}_{\mu_g}& \tilde{\bm{\Sigma}}^{-1}_{\mu_g\beta_g}\\
 \tilde{\bm{\Sigma}}^{-1}_{\mu_g\beta_g},&\tilde{\bm{\Sigma}}^{-1}_{\beta_g}
\end{array}\right)^{-1} \right],
\end{equation*}
 where  \begin{eqnarray*}
\tilde{\bm{\Sigma}}^{-1}_{\mu_g}&=a_{g,4}\tilde{\bm{\Sigma}}_g^{-1},\qquad&\qquad
 \bar{\Mu}_g=\frac{a_{g,3}}{(a_{g,3}a_{g,4}-a_{g,0}^2)}\left(\mathbf{a}_{g,2}-\frac{\mathbf{a}_{g,1}a_{g,0}}{a_{g,3}}\right),\\
\tilde{\bm{\Sigma}}^{-1}_{\beta_g}&=a_{g,3} \tilde{\bm{\Sigma}}_g^{-1},\qquad&\qquad
 \bar{\Beta}_g=\frac{a_{g,4}}{(a_{g,3}a_{g,4}-a_{g,0}^2)}\left(\mathbf{a}_{g,1}-\frac{\mathbf{a}_{g,2}a_{g,0}}{a_{g,4}}\right),\\
 \tilde{\bm{\Sigma}}^{-1}_{\mu_g\beta_g}&=a_{g,0} \tilde{\bm{\Sigma}}_g^{-1}.&\\
 \end{eqnarray*}
 
 A truncated normal prior distribution was assigned to $\tilde{\gamma}_g$ such that $$\tilde{\gamma}_g \sim \text{N}({a_{g,0}^{(0)}}/{a_{g,3}^{(0)}}, {1}/{2a_{g,3}^{(0)}}) \text{I}(\tilde{\gamma}_g>0),$$ 
and so the posterior distribution for $\tilde{\gamma}_g$ is given by $$\tilde{\gamma}_g \sim \text{N}({a_{g,0}}/{a_{g,3}}, {1}/{2a_{g,3}}) \text{I}(\tilde{\gamma}_g>0).$$ 

For the MNIG model,
\begin{equation*}\begin{split}
\ex[\log~p(\by_i,z_{ig}=1,u_{ig}&\mid\boldsymbol{\theta})]_{q_{\btheta}(\boldsymbol{\theta})}=\ex[\log(\pi_g)]-\frac{d+1}{2}\log(2\pi)-\frac{d+3}{2} \log (u_{ig})\\&+\frac{1}{2}\ex[\log |\tilde{\bm{\Sigma}}_g^{-1}|)]+\ex[\tilde{\gamma}_g]-\frac{1}{2}\left[\ex[A_{ig}]u_{ig}^{-1}+\ex[B_g]u_{ig}\right],\\
\end{split}\end{equation*}
where $A_{ig}=1+(\mathbf{y}_i -\tilde{\boldsymbol{\mu}}_g)'\tilde{\bm{\Sigma}}_g^{-1}(\mathbf{y}_i -\tilde{\Mu}_g)$, $B_g=\tilde{\gamma}_g^2+\tilde{\Beta}_g\tilde{\bm{\Sigma}}_g^{-1}\tilde{\Beta}_g'$, and $C_{ig}=\tilde{\gamma}_g+(\by_i-\tilde{\Mu}_g)'\tilde{\bm{\Sigma}}_g^{-1}\tilde{\Beta}_g$.

Following \cite{beal2003}, the approximating joint density of missing variables $(z_{ig},u_{ig})$ for the conjugate-exponential models can be obtained as
\begin{equation*}\begin{split}
q_{z_{ig},u_{ig}}(z_{ig}=1,u_{ig})\propto&\exp\left\{\ex[\log(\pi_g)]-\frac{d+1}{2}\log(2\pi)-\frac{d+3}{2} \log (u_{ig})+\frac{1}{2}\ex[\log |\tilde{\bm{\Sigma}}_g^{-1}|)]\right.\\
 &\qquad\qquad\qquad\qquad+ \left.\ex[C_{ig}]-\frac{1}{2}\left[\ex[A_{ig}]u_{ig}^{-1}+\ex[B_g]u_{ig}\right]\right\}\\
 &=(2\pi)^{-\frac{d+1}{2}}\exp\left\{\ex[\log(\pi_g)]+\frac{1}{2}\ex[\log |\tilde{\bm{\Sigma}}_g^{-1}|)]+\ex[C_{ig}]\right\}\\
 &\qquad\qquad\qquad\qquad\times u_{ig}^{-\frac{d+3}{2}}\exp\left\{-\frac{1}{2}\left[\ex[A_{ig}]u_{ig}^{-1}+\ex[B_g]u_{ig}\right]\right\}\\
 &=(2\pi)^{-\frac{d+1}{2}}\exp\left\{\ex[\log(\pi_g)]+\frac{1}{2}\ex[\log |\tilde{\bm{\Sigma}}_g^{-1}|)]+\ex[C_{ig}]\right\}\\ &\qquad\qquad\qquad\qquad\times\text{GIG}\left(u_{ig}~\bigg|~-\frac{d+1}{2},\sqrt{\ex[A_{ig}]},\sqrt{\ex[B_g]}\right).
\end{split}\end{equation*}
 Here, $\text{GIG}(\cdot)$ is the probability density function of the generalized inverse Gaussian distribution and 
 \begin{eqnarray*}
 \ex[\log(\pi_g)]&=&\Psi(a_{g,0})-\Psi(n) ,\\
 \ex[\log |\tilde{\bm{\Sigma}}_g^{-1}|]&=&\sum_{s=1}^d\Psi\left(\frac{a_{g,0}+1-s}{2}\right)+d \log(2)-\log|\mathbf{V}_g|,\\
 \ex[A_{ig}]&=&\ex[1+(\mathbf{y}_i -\tilde{\boldsymbol{\mu}}_g)'\tilde{\bm{\Sigma}}_g^{-1}(\mathbf{y}_i -\tilde{\Mu}_g)]\\
 &=&1+(\mathbf{y}_i -\ex[\tilde{\boldsymbol{\mu}}_g])'\ex[\tilde{\bm{\Sigma}}^{-1}_g](\mathbf{y}_i -\ex[\tilde{\Mu}_g])+\tr\left\{\ex[\tilde{\bm{\Sigma}}^{-1}_g]\text{Var}(\tilde{\Mu}_g)\right\},\\
 \ex[B_g]&=&\ex[\tilde{\gamma}_g^2+\tilde{\Beta}_g\tilde{\bm{\Sigma}}_g^{-1}\tilde{\Beta}_g']\\
 &=&(\ex[\tilde{\gamma}_g])^2+\text{Var}(\tilde{\gamma}_g)+\ex[\tilde{\Beta}_g]\ex[\tilde{\bm{\Sigma}}_g^{-1}]\ex[\tilde{\Beta}_g]+\tr\left\{\ex[\tilde{\bm{\Sigma}}^{-1}]\text{Var}(\tilde{\Beta}_g)\right\},\\
  \ex[C_{ig}]&=&\ex[\tilde{\gamma}_g+(\by_i-\tilde{\Mu}_g)'\tilde{\bm{\Sigma}}_g^{-1}\tilde{\Beta}_g]\\
 &=&\ex[\tilde{\gamma}_g]+\by_i\ex[\tilde{\bm{\Sigma}}_g^{-1}]\ex[\tilde{\Beta}_g]-\ex[\tilde{\Mu}'_g]\ex[\tilde{\bm{\Sigma}}_g^{-1}]\ex[\tilde{\Beta}_g]
+\tr\left\{\ex[\tilde{\bm{\Sigma}}^{-1}]\text{Cov}(\tilde{\Mu}_g,\tilde{\Beta}_g)\right\},
 \end{eqnarray*}
 where $\Psi(\cdot)$ is the digamma function.
 
 The approximating density $q_{z_{ig}}(z_{ig}=1)$ is
\begin{equation*}\begin{split}
   q_{z_{ig}}(z_{ig}=1)&=\int_{u_{ig}} q_{z_{ig},u_{ig}}(z_{ig}=1,u_{ig}) du_{ig}\\
   &\propto\int_{u_{ig}}(2\pi)^{-\frac{d+1}{2}}\exp \left\{\ex[\log(\pi_g)]+\frac{1}{2}\ex[\log |\tilde{\bm{\Sigma}}_g^{-1}|]+\ex[C_{ig}]\right\}\\
   &\qquad\qquad\qquad\quad\times\text{GIG}\left(u_{ig}~\bigg|~-\frac{d+1}{2},\sqrt{\ex[A_{ig}]},\sqrt{\ex[B_g]}\right)du_{ig}\\
   &=(2\pi)^{-\frac{d+1}{2}}\exp\left\{\ex[\log(\pi_g)]+\frac{1}{2}\ex[\log |\tilde{\bm{\Sigma}}_g^{-1}|]+\ex[C_{ig}]\right\}\\
   &\qquad\qquad\qquad\quad\times 2\left(\frac{\ex[A_{ig}]}{\ex[B_g]}\right)^{-\frac{d+1}{2}}K_{-\frac{d+1}{2}}\left(\sqrt{\ex[A_{ig}]\ex[B_g]}\right).
\end{split}\end{equation*}
The probability that $z_{ig}=1$ is
 $$\hat{z}_{ig}=\frac{q_{z_{ig}}(z_{ig}=1)}{\sum_{g=1}^Gq_{z_{ig}}(z_{ig}=1)}.$$
The density $q_{u_{ig}}(u_{ig}\mid z_{ig}=1)$ is
\begin{eqnarray*}
q_{u_{ig}}(u_{ig}\mid z_{ig}=1)&\propto& (2\pi)^{-\frac{d+1}{2}}\exp\left\{\ex[\log(\pi_g)]+\frac{1}{2}\ex[\log |\tilde{\bm{\Sigma}}_g^{-1}|]+\ex[C_{ig}]\right\}\\
&&\qquad\times 2\left(\frac{\ex[A_{ig}]}{\ex[B_g]}\right)^{-(d+1)/2}K_{-\frac{d+1}{2}}\left(\sqrt{\ex[A_{ig}]\ex[B_g]}\right),
\end{eqnarray*}
and so $U_{ig}\mid({z}_{ig}=1)\backsim\text{GIG}(-\frac{d+1}{2},\sqrt{\ex[A_{ig}]},\sqrt{\ex[B_g]}).$ Therefore,
\begin{eqnarray*}
\ex[U_{ig}|z_{ig}=1]_{q_{u_{ig}}(u_{ig}\mid z_{ig}=1)}&=&\left(\frac{\ex[A_{ig}]}{\ex[B_g]}\right)^{\frac{d+1}{2}}\frac{K_{-\frac{d-1}{2}}(\sqrt{\ex[A_{ig}]\ex[B_g]})}{K_{-\frac{d+1}{2}}(\sqrt{\ex[A_{ig}]\ex[B_g]})},\\
\ex[U^{-1}_{ig}|z_{ig}=1]_{q_{u_{ig}}(u_{ig}\mid z_{ig}=1)}&=&\left(\frac{\ex[A_{ig}]}{\ex[B_g]}\right)^{-\frac{d+1}{2}}\frac{K_{-\frac{d+3}{2}}(\sqrt{\ex[A_{ig}]\ex[B_g]})}{K_{-\frac{d+1}{2}}(\sqrt{\ex[A_{ig}]\ex[B_g]})}.
\end{eqnarray*}

Similar to the univariate approach, the variational Bayes algorithm proceeds in the following manner: 
 \begin{itemize}
 \item For the observed data $\by=(\by_1,\by_2,\ldots,\by_n)$, the algorithm is initialized with more components than expected, say $G$. The $\hat{z}_{ig}$ can be initialized by either randomly assigning the observations to one of the $G$ components or by using the results from another clustering method (e.g., $k$-means clustering). 
\item Using the initialized values of $\hat{z}_{ig}$, the parameter of the $g$th component is initialized as follows:
\begin{itemize}
\item The component's sample mean is used to initialize the parameter $\Mu_g$, 
\item the component's sample covariance is used to initialize $\boldsymbol{\Sigma}_g$, 
\item $\bbeta$ is set to $\mathbf{0}$, and 
\item $\gamma$ is set to 1. 
\end{itemize}
\item Using these values of the parameters, the expected values of  $U^{-1}_{ig}$ and $U_{ig}$ are initialized.
\item The hyperparameters of the prior distribution are initialized to have a flat distribution over the possible values of the parameters.
\begin{enumerate}
\item Using the $\hat{z}_{ig}$ and expected values of $U^{-1}_{ig}$ and $U_{ig}$, the hyperparameters of the approximating density $q_\theta(\btheta)$ are updated. Using these updated hyperparameters, the expected values $\ex[\log r(\btheta)]$ and $\ex[\phi_j(\btheta)]$ are updated.\label{step1}
\item Using these updated $\ex[\log r(\btheta)]$ and $\ex[\phi_j(\btheta)]$, the $\hat{z}_{ig}$, $\ex[U^{-1}_{ig}|z_{ig}=1]$, and $\ex[U_{ig}|z_{ig}=1]$ are updated.\label{step2}
\item Components with too few observations are eliminated. Specifically, for each component $g'$ we do the following. If the estimated number of observations in component $g'$, i.e., $\sum_{i=1}^n\hat{z}_{ig'}$, is sufficiently small (less than one in our case), then component $g'$ is eliminated.\label{step3}
\end{enumerate}
Steps \ref{step1}, \ref{step2}, and  \ref{step3} are repeated until convergence.
\end{itemize}

As in the univariate case, after convergence is achieved, the observations are assigned to components using the MAP.
%As in the univariate case, once convergence is achieved, the observations were assigned to components using maximum \textit{a~posteriori} probability (MAP).

\subsection{Simulated Data}
\subsubsection{Simulation Study 4}
\begin{figure}[!htbp]
\vspace{-0.2in}
\includegraphics[width=0.8\textwidth]{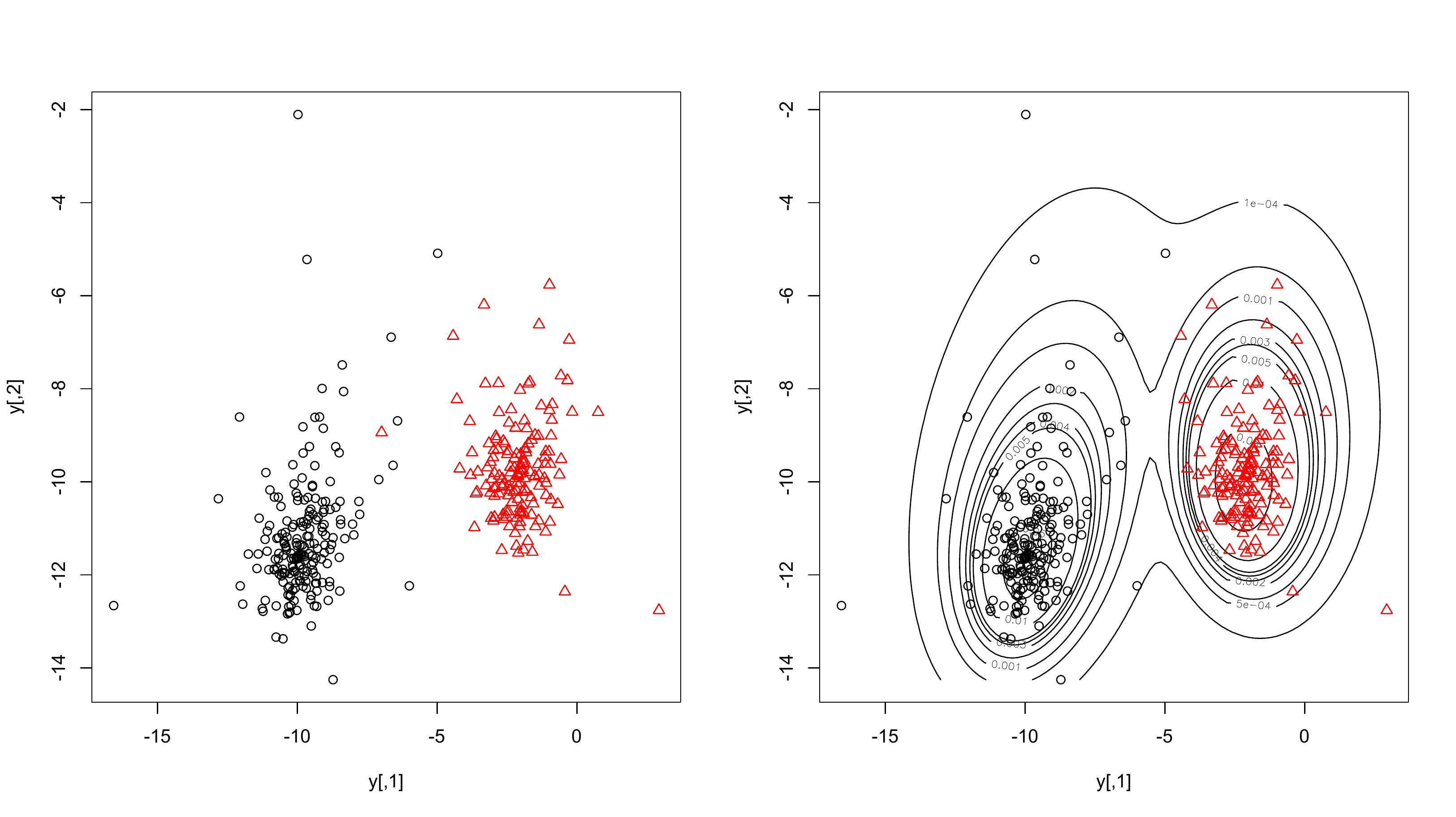}
\caption{Scatter plot highlighting the true labels for the simulated data from Simulation Study~4 (left) and  a contour plot showing the predicted classifications (right).}
\label{fig3}
\end{figure}

To demonstrate the recovery of underlying parameters, our variational Bayes algorithm was applied to a simulated two-dimensional data set (Figure~\ref{fig3}) with two symmetric components ($n_1=150$ and $n_2=200$). Our algorithm was initialized with $G=5$ components and, after running to convergence, gave a two-component model with one misclassified observation ($\text{ARI}=0.99$). The estimated parameters are very close to the true values, as can be seen in Table~\ref{tab2} and below: 
\begin{eqnarray*}
\tilde{\bm{\Sigma}}_1 &=& \left(
\begin{array}{cc}
1.2 & 0 \\
0& 1.2 
\end{array} \right), \ \hat{\tilde{\bm{\Sigma}}}_1 = \left(
\begin{array}{cc}
0.75& 0.05 \\
0.05& 0.68
\end{array} \right); \
 \tilde{\bm{\Sigma}}_2 = \left(
\begin{array}{cc}
1 & 0.4 \\
0.4& 1 
\end{array} \right), \ \hat{\tilde{\bm{\Sigma}}}_2 = \left(
\begin{array}{cc}
1.32 & -0.35\\
-0.35& 1.28
\end{array} \right). \label{e2}
\end{eqnarray*}
\begin{table}[!htbp]
%\vspace{-0.2in}
\caption{Estimated and true values for the parameters of the MNIG model in Simulation Study~4.}
\begin{tabular*}{1.0\textwidth}{@{\extracolsep{\fill}}llrrr}
\hline
& &$\tilde{\Beta}$ & $\tilde{\gamma}$ &$\tilde{\Mu}$\\
\hline
\multirow{2}{*}{Component 1}& True&$(0.1,0.2)$&$1.2$&$(-2,-10)$\\
&Estimated&$(0.17,0.49)$&$1.63$&$(-2.39,-10.35)$\\
\hline
\multirow{2}{*}{Component 2}& True&($0.2$,$0.75$)&$0.8$&($-10$,$-12$)\\
& Estimated&$(0.53,1.00)$&$0.69$&$(-10.00,-11.89)$\\
\hline
\end{tabular*}
\label{tab2}
\end{table}
\FloatBarrier
Our model clearly fits the data very well, with the contours capturing the shape of the two components (Figure~\ref{fig3}). The parameter estimates must be considered in context with the actual fit of the model (Figure~\ref{fig3}) because it is known that different parameter sets can give very similar densities for these models \citep{Lillestol2000}.
\FloatBarrier

\subsubsection{Simulation Study 5}\label{multisimex2}

\begin{figure}[htb]
\includegraphics[width=\textwidth]{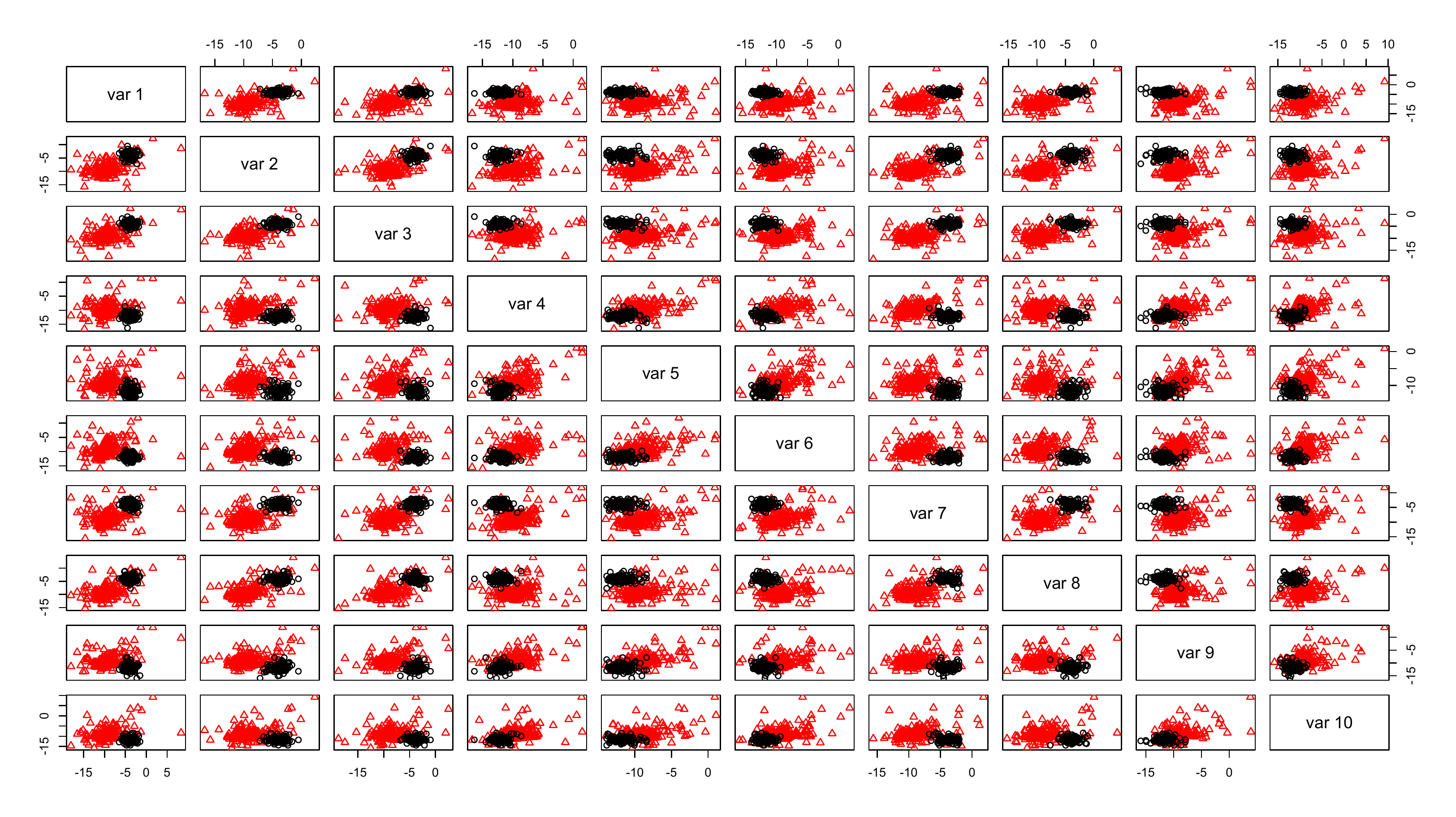}
\caption{Pairs plot showing the true classifications for the data from Simulation~5.}
\label{figsim-m3}
\end{figure}

To present a more challenging and higher dimensional example, we generated a ten-dimensional data set with two components ($n_1=150$ and $n_2=200$) that are not well separated (Figure~\ref{figsim-m3}). The variational Bayes algorithm was run starting with $G = 10$ components, resulting in a two-component model with perfect classification ($\text{ARI}=1$). %Although not shown here, the estimated parameters were close to the true values. 
 
\subsection{Old Faithful Data}
\begin{figure}[!h]
\vspace{-0.3in}
\includegraphics[width=0.55\textwidth]{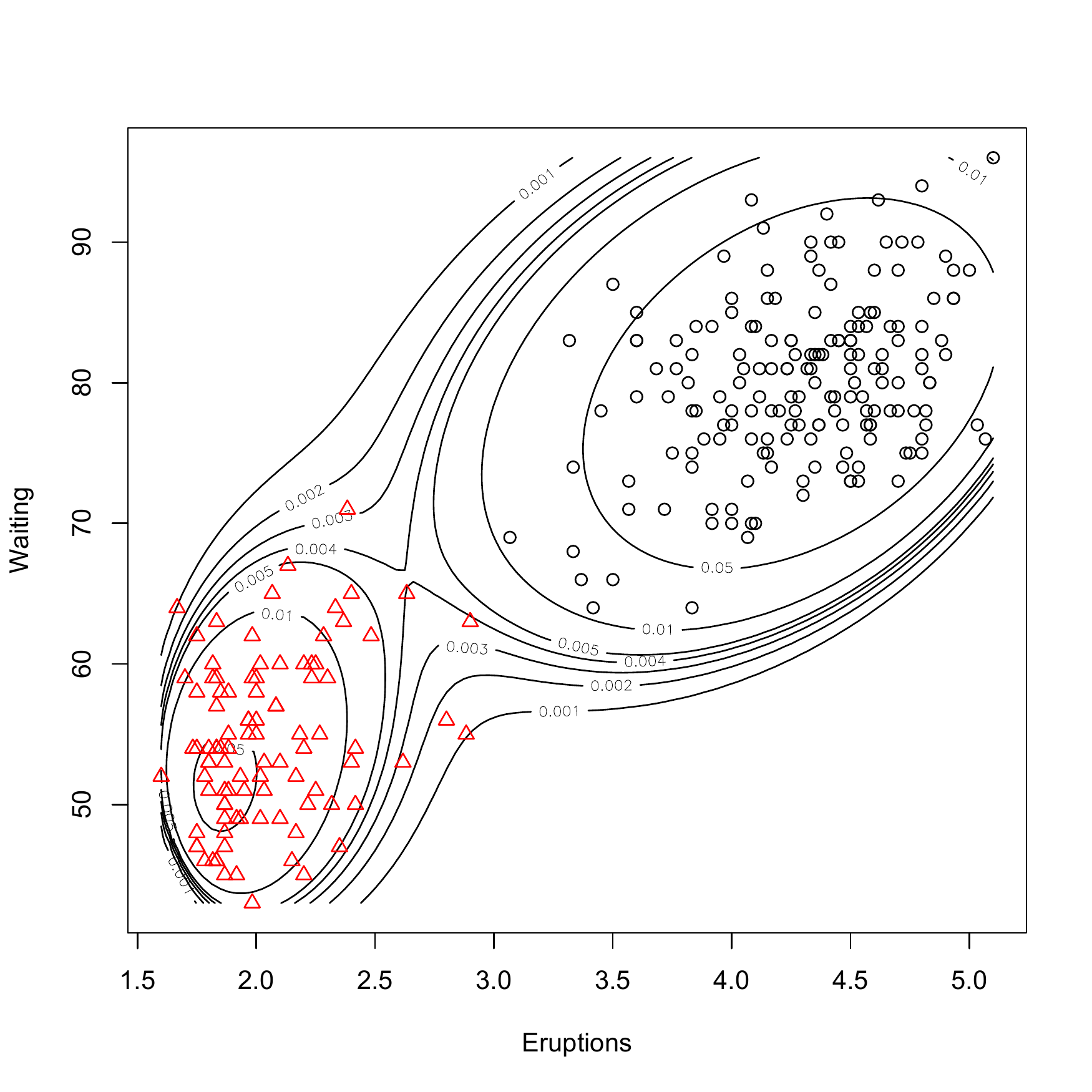}
\caption{Contour plot of the Old Faithful data using the fitted two-component mixture model.}
\label{figoldfaithful}
\end{figure}

The Old Faithful data are available in the {\sf R} package \texttt{MASS} \citep{mass2002}. These data comprise the waiting time between and the duration of 272 eruptions of the Old Faithful geyser in Yellowstone National Park, Wyoming, USA. These data do not contain true labels; however, upon visual inspection, the data seem to have two classes: shorter, more frequent eruptions and longer, less frequent eruptions.
We ran our variational Bayes algorithm on the Old Faithful data starting with $G=7$ components. The resulting $G=2$ component model fits the data very well (Figure~\ref{figoldfaithful}).

Several others have used non-Gaussian model-based clustering on these data, via variants of the EM algorithm, and obtained similar results \citep[e.g.,][]{franczak2012,vrbik12}.

\subsection{Crabs data} 
\begin{figure}[htbp]
 \vspace{-0.4in}
\hspace{-0.1in}
\includegraphics[width=0.535\textwidth,height=2.25in]{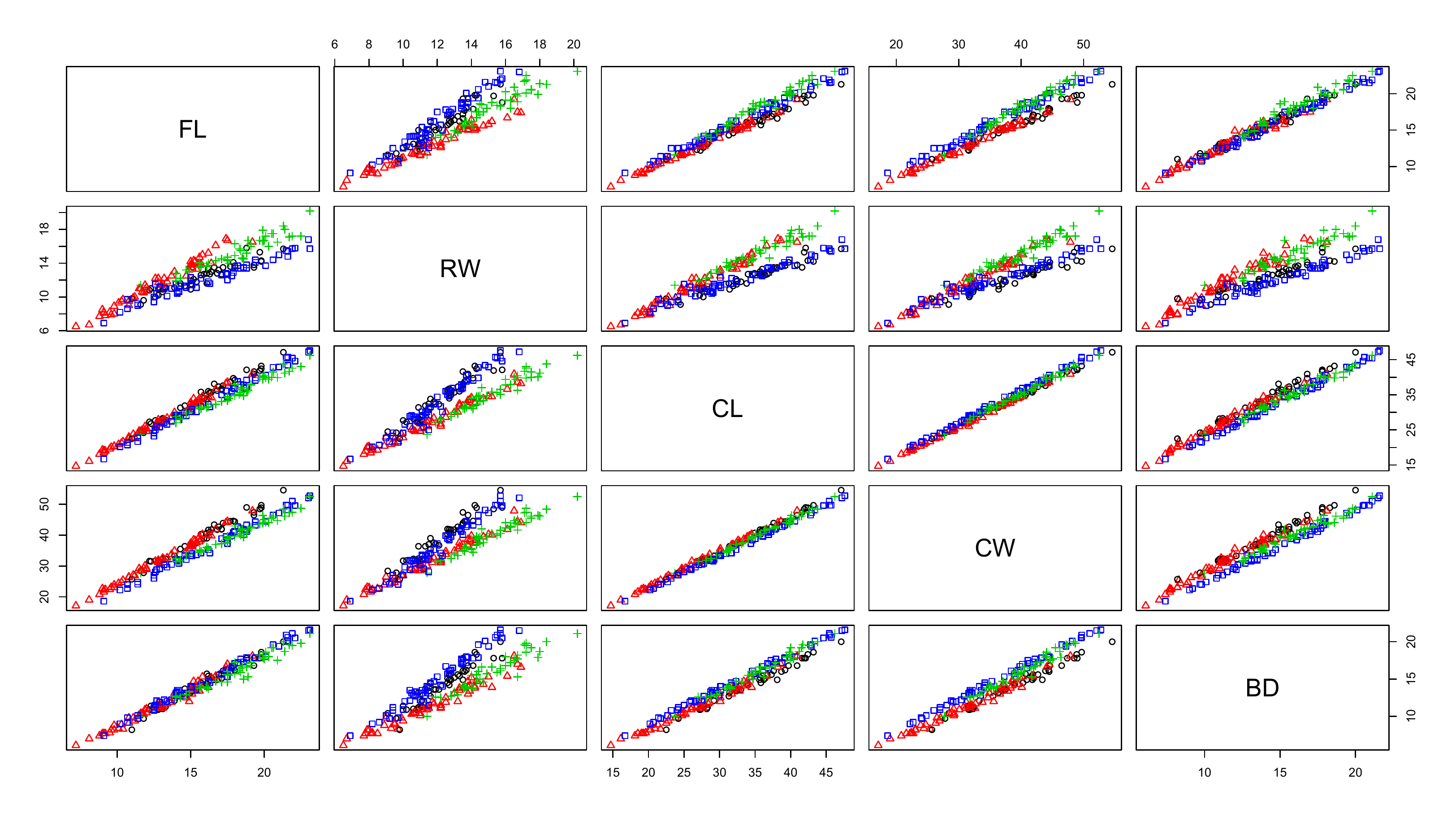}
\hspace{-0.15in}
\includegraphics[width=0.535\textwidth,height=2.25in]{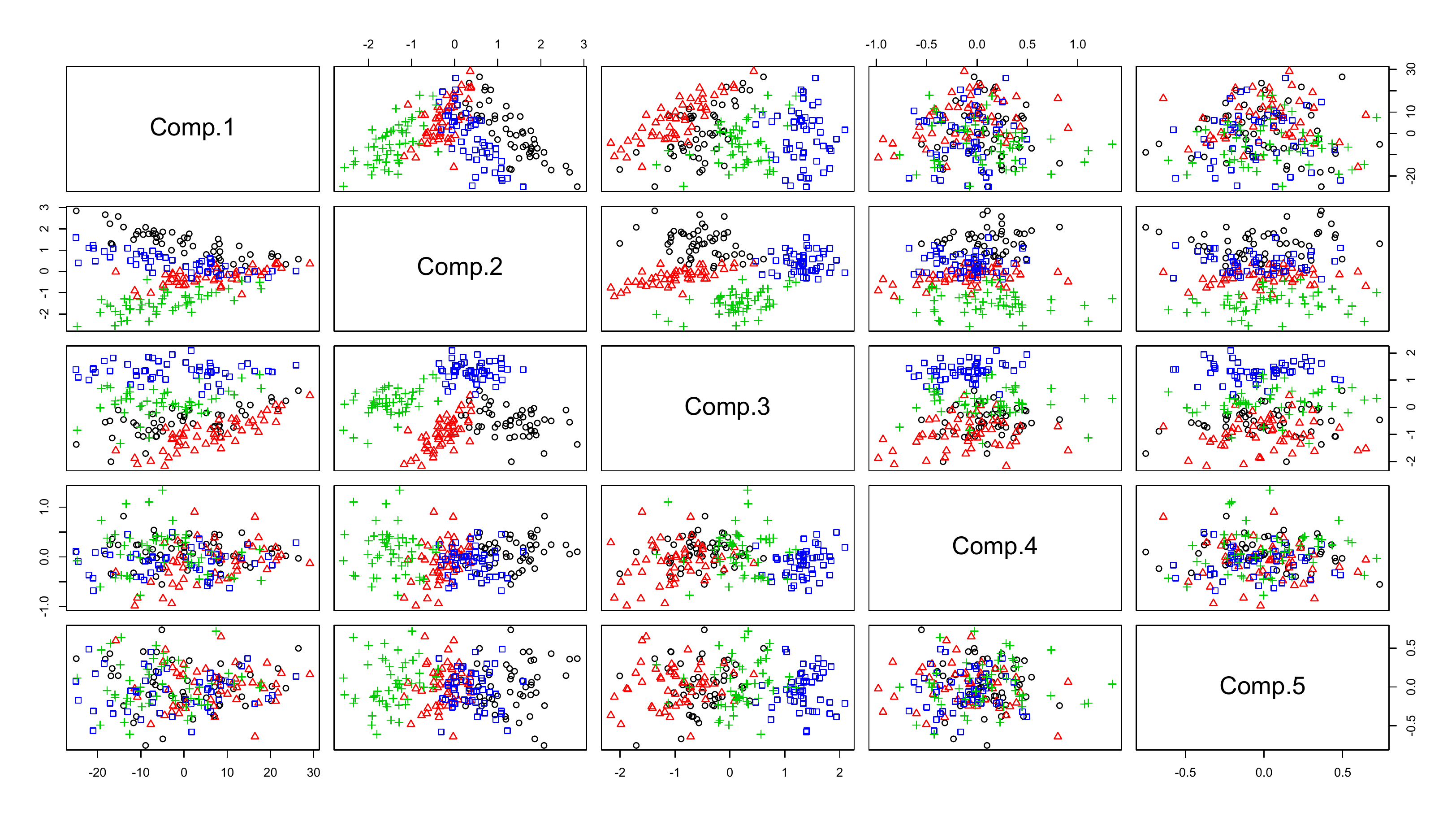}
\caption{Pairs plots showing estimated classifications for the crabs data using the variables (left) and the principal component (right), respectively.}
\label{figestcrabs}
\end{figure}

The crabs data, available in the {\sf R} package \texttt{MASS}  \citep{mass2002}, contain morphological measurements of 50 male and 50 female crabs (\textit{Leptograpsus variegatus}) in each of the two colour forms: blue and orange. The measured morphological variables are frontal lobe size (FL), rear width (RW), carapace length (CL), carapace width (CW), and body depth (BD), all measured in mm.
%\begin{figure}[htbp]
%\includegraphics[width=0.8\textwidth]{crabs-true.pdf}
%\caption{Matrix scatter plot of all variables in the crab data set.}
%\label{figtruecrabs}
%\end{figure}
% 
We ran our variational Bayes algorithm on the crab data starting with $G=10$ components. This resulted in a $G=4$ component model with an associated ARI of 0.79 (Table~\ref{tabcrabs}). The variables in these data are highly correlated and so visualizing classification performance is easier in the principal component space (Figure~\ref{figestcrabs}). 
 \begin{table}[htbp]
\begin{center}
\caption{The new merged ``true" classification and the estimated classification.}
\begin{tabular*}{1.0\textwidth}{@{\extracolsep{\fill}}l|llll}
\hline
 True $\backslash$ Estimates&1&2&3&4\\
\hline
  Blue\&Male &41  &9& & \\
  Blue\&Female&1&  48&&1   \\
Orange\&Male  && &  50&  \\
  Orange\&Female &&  &6  &  44\\
\hline
\end{tabular*}
\label{tabcrabs}
\end{center}
\end{table}

Because the variables in these data are so highly correlated, they are very difficult to cluster. In fact, they are notoriously difficult to cluster and our variational Bayes MNIG performs very well when compared to many other approaches. To find methods in the literature that produce a higher ARI than $0.79$, one needs to look at families of mixture models \citep[e.g.,][]{mcnicholas10a,andrews12,vrbik13} or methods that combine variable selection and clustering \cite[e.g.,][]{morris13b}. Of course, if our variational Bayes MNIG approach was extended to incorporate a family of models, an ARI above $0.79$ might be achieved (cf.\ Section~\ref{sec:4}). 

\subsection{Fish Catch Data}
The fish catch data, available in the {\sf R} package \texttt{rrcov} \citep{rrcov09}, contain different measurements on the body size and weight of seven different fish species (bream, whitewish, roach, parkki, smelt, pike, and perch). The variable \texttt{Weight} gives the weight of the fish in grams, \texttt{Length1} is the length from the nose to the beginning of the tail, \texttt{Length2}  is the length from the nose to the notch of the tail, \texttt{Length3} is the length from the nose to the end of the tail, \texttt{Height} is the maximal height as a percentage of \texttt{Length3}, and \texttt{Width} is maximal width as a percentage of \texttt{Length3}.

\begin{figure}[htbp]
\vspace{-0.2in}
\hspace{-0.1in}
\includegraphics[width=\textwidth]{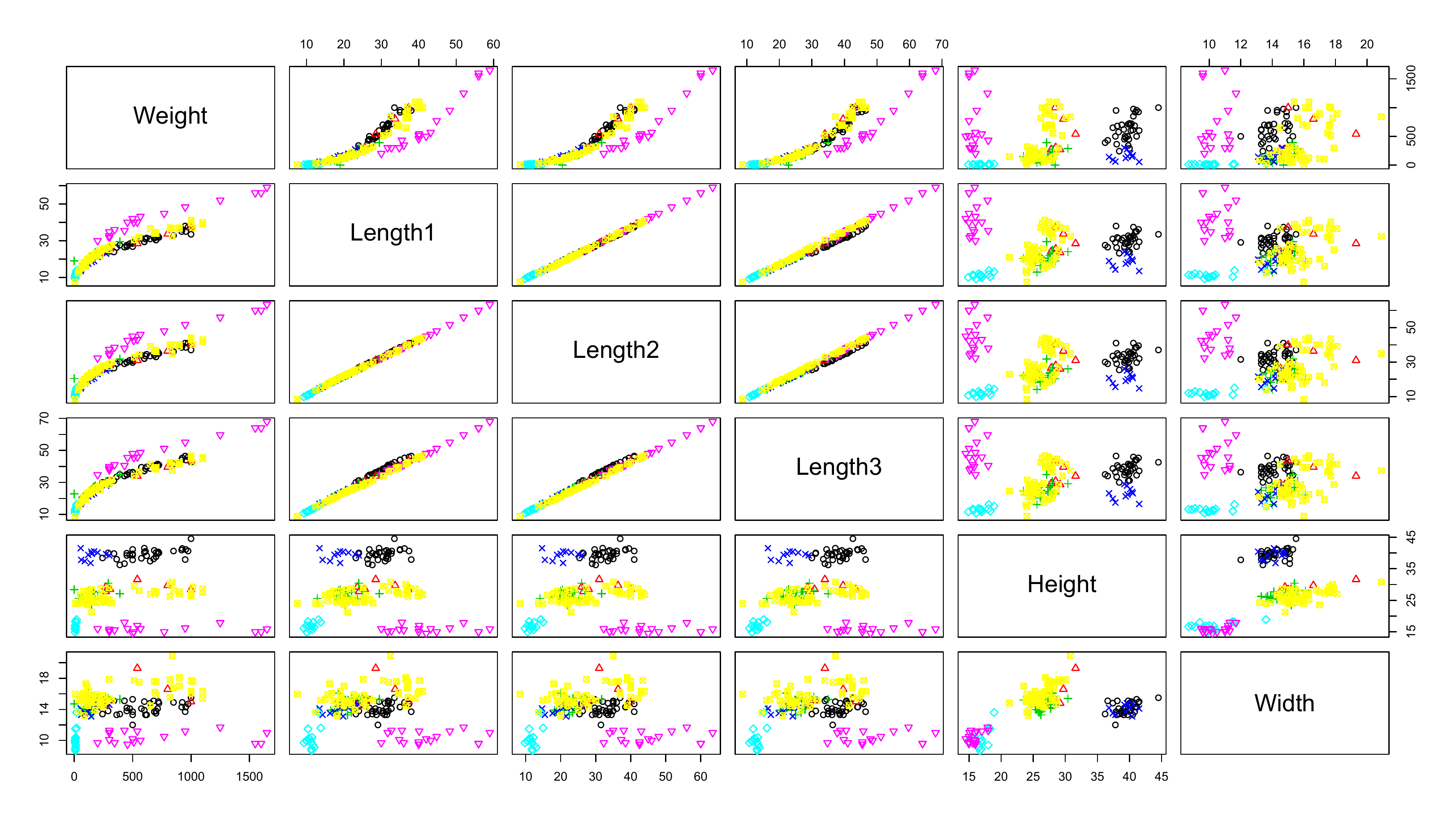}
\caption{Matrix scatter plot of all variables in the fish catch data set, where different colours represent different species.}
\label{fig4}
\end{figure}

As expected, all of the length measurements are very highly correlated with each other (correlation $>$ 0.99) and with the weight measurements (correlation $>$ 0.91), cf.\ Figure~\ref{fig4}. Therefore, the highly correlated variables \texttt{Length1}, \texttt{Length2}, and \texttt{Weight} were dropped from further analysis. These data were explored by \cite{karlis09}, who dropped \texttt{Length1}, \texttt{Length2}, and \texttt{Height} before their analysis.

\begin{figure}[htbp]
\vspace{-0.1in}
\hspace{-0.075in}\includegraphics[width=0.5325\textwidth,height=2.5in]{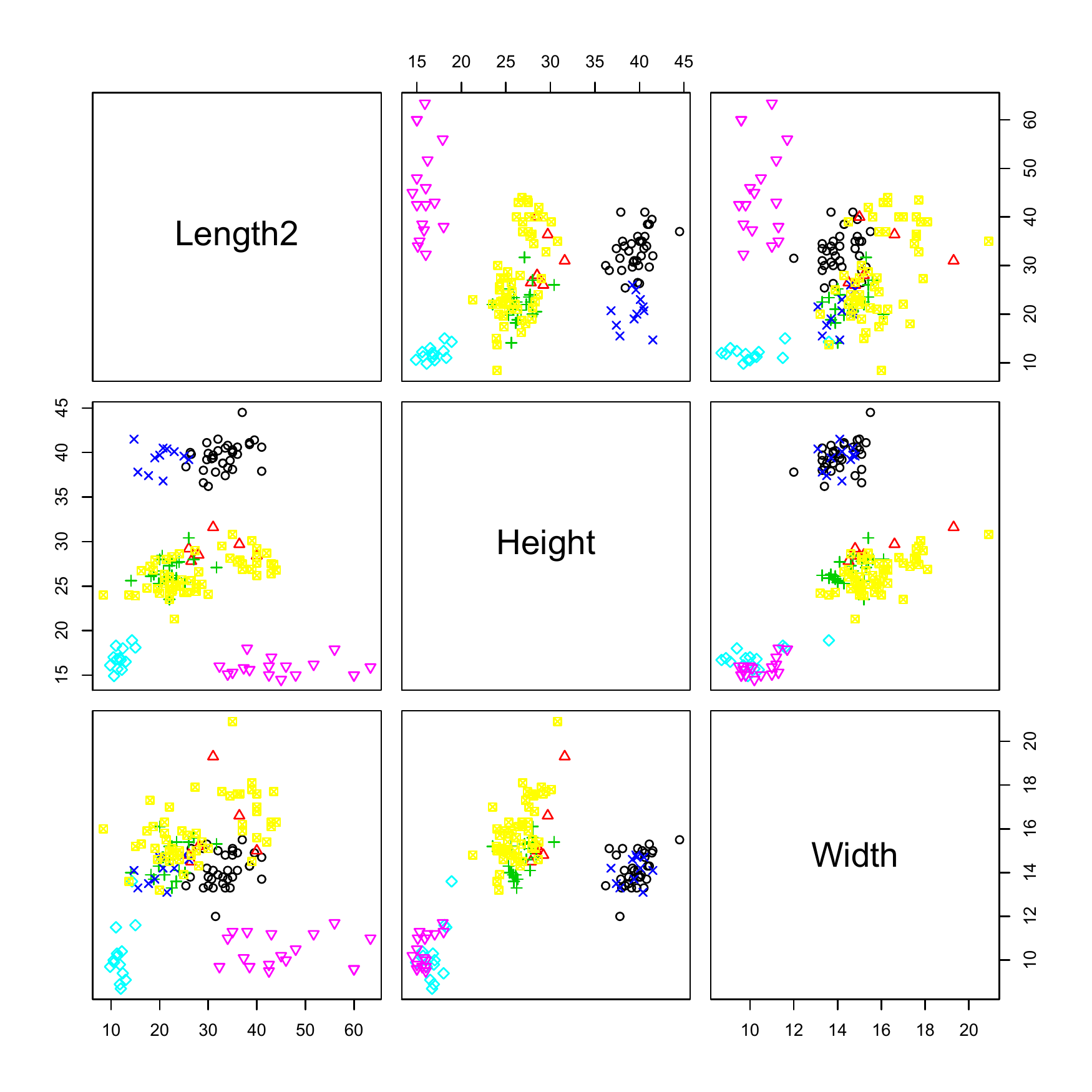}
\hspace{-0.125in}\includegraphics[width=0.5325\textwidth,height=2.5in]{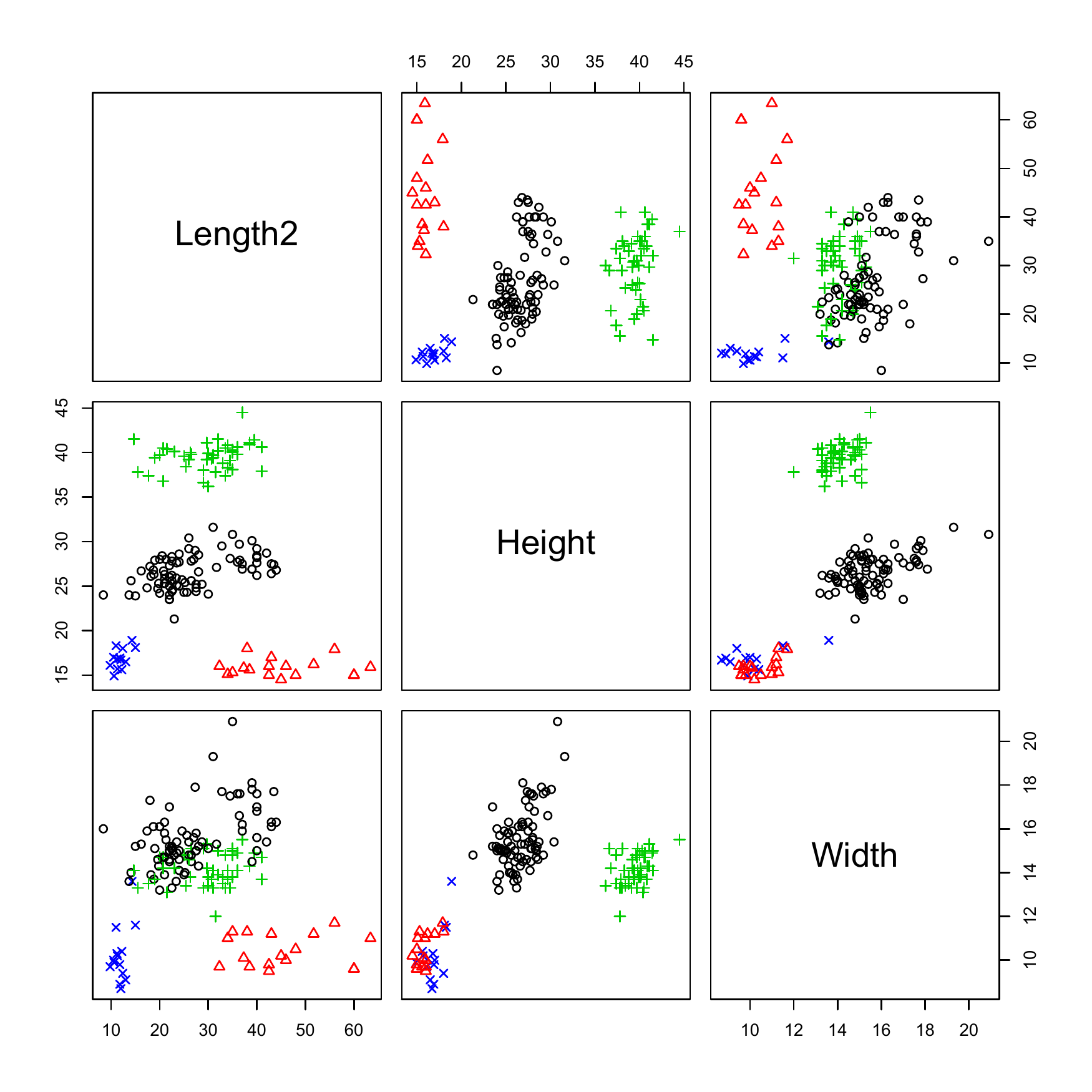}
\caption{True (left) and estimated (right) classifications for the fish catch data based on the variables \texttt{Length2}, \texttt{Height}, and \texttt{Width}.}
\label{fig5}
\end{figure}

We ran our variational Bayes algorithm on the resulting three-dimensional data set, starting with $G=10$ components. This resulted in a $G=4$ component model with the classifications shown in Figure~\ref{fig5}. %corresponding to an ARI of 0.63. 
By inspection of Figure~\ref{fig5}, we can see that a four-component solution is not unreasonable based on the three variables used (\texttt{Length2}, \texttt{Height}, and \texttt{Weight}). Classification results for our four-component model (Table~\ref{tab3}) correspond exactly to a merging of Species~2, 3, and~7 (whitewish, roach, and perch), and Species~1 and~4 (bream and parkki). \cite{karlis09}  who used different variables --- \texttt{Length3}, \texttt{Weight}, and \texttt{Width} --- in an EM framework, obtained a seven-component model using the Akaike information criterion \citep[AIC;][]{akaike73} for model selection and a four-component model using the Bayesian information criterion \citep[BIC;][]{schwartz78}.
\begin{table}[htbp]
\begin{center}
\caption{Cross-tabulation of true versus predicted classifications for the fish catch data.}
\begin{tabular*}{1.0\textwidth}{@{\extracolsep{\fill}}l|llll}
\hline
True$\backslash$ Estimates&1&2&3&4\\
\hline
\texttt{Bream}  &34  && & \\
\texttt{Parkki}  &11  && & \\
\texttt{Whitewish} &&  6& &  \\
\texttt{Roach}&&  20& &  \\
 \texttt{Perch}  &&  56& &  \\
\texttt{Smelt}  && &  14&  \\
\texttt{Pike} &&  &  &  17\\
\hline
\end{tabular*}
\label{tab3}
\end{center}
\end{table}
%\begin{table}[htbp]
%\begin{center}
%\caption{The new merged ``true" classification and the estimated classification.}
%\begin{tabular*}{1.0\textwidth}{@{\extracolsep{\fill}}l|llll}
%\hline
%True$\backslash$ Estimates&1&2&3&4\\
%\hline
%\texttt{Bream} and \texttt{Parkki}  &45  && & \\
%\texttt{Whitewish, Roach}, and \texttt{Perch}  &&  82& &  \\
%\texttt{Smelt}  && &  14&  \\
%\texttt{Pike} &&  &  &  17\\
%\hline
%\end{tabular*}
%\label{tab3}
%\end{center}
%\end{table}

 \section{Conclusion}
 \label{sec:4}
Variational Bayes approximations are presented as an effective alternative to the EM
algorithm for parameter estimation for UNIG and MNIG mixtures. They have been
used for Gaussian mixture models in the past; however, this is their first application for non-Gaussian models. Furthermore, it is the first time variational approximations have been used for non-symmetric distributions. Although we illustrated our variational Bayes approach through model-based clustering, it could be applied to model-based classification \citep[e.g.,][]{mcnicholas10c} or discriminant analysis \citep{hastie96} in an analogous fashion. In this paper, we illustrate that variational Bayes approximations can be very effective for non-Gaussian mixture model-based clustering, classification, and discriminant analysis. Accordingly, this paper may well be the forerunner to several others detailing the application of variational Bayes approximations in the complex modelling situations that can arise in non-Gaussian model-based clustering.

As reported by \cite{karlis09}, the EM algorithm for MNIG takes a very long time to converge. Therefore, running multiple EM algorithms to cover a range of values for $G$, which is needed when the ÔtrueÕ number of components is unknown, adds to an already heavy computational burden. Variational Bayes approximations, on the other hand, start off with more components than expected, and once the number of observations in a component becomes sufficiently small, it is removed. This allows for simultaneous parameter estimation and estimation of the number of components, and is far more computationally efficient than running an EM algorithm for each of several possible values of $G$.

We demonstrated the efficacy of our approach by clustering real and simulated data for both the UNIG and MNIG mixtures. Some possible avenues for further research include extending these models, and the associated parameter estimation approach, to achieve parsimony. The could be carried out via imposing constraints upon an eigen-decomposition of the component scale matrices by analogy with the work of \cite{celeux95} on Gaussian mixtures. Our variational approximations could be extended to mixtures of factor analyzers \citep{ghahramani97,mclachlan00a}, or a variatiants thereof \citep{mcnicholas08,mcnicholas10b,Baek2010,murray13b}, where it may be possible to select the number of latent factors in addition to the number of components. Applications aimed at longitudinal data analysis \citep[e.g.][]{mcnicholas10b,mcnicholas12} and contaminated mixtures \citep{punzo13} will also be considered within a variational framework.

\section*{Acknowledgements}
This work was supported by a Postgraduate Scholarship from the Natural Sciences and Engineering Research Council of Canada, an Early Researcher Award from the Ontario Ministry of Research and Innovation, and the University Research Chair in Computational Statistics.

%\bibliographystyle{chicago}      % basic style, author-year citations
%\bibliography{sanjeenasbib}   % name your BibTeX data base

\end{document}